\documentclass[prd, aps, reprint, amsfonts, amssymb, amsmath, showpacs, nofootinbib, superscriptaddress, preprintnumbers, ]{revtex4-1}
\usepackage[english]{babel}
\usepackage[utf8]{inputenc}
\usepackage{myterms}
\usepackage{amsmath}	%Allows \begin{equation}, \begin{align}, \underbrace{2+2=4}_{\text{obvious}}
\usepackage{mathrsfs}     %Allows \mathscr{L}
\usepackage{bm}		%Allows {\bm \nabla \alpha} for bold symbold
\usepackage{amssymb}	%Allows use math symbols
\usepackage{color} 		%Allows {\color{red} Hello World} or \colorbox{blue}{Hello World}
\usepackage{graphicx} 	%Allows \includegraphics{figure.pdf}
\usepackage{times} 		%Use Times New Roman font
\usepackage[T1]{fontenc}
\usepackage{mathrsfs} %Allows \mathscr{HELLO}
\usepackage{multirow} 	% Allows \multirow{3}{*}{XXXX}
\usepackage[citecolor=blue]{hyperref}\usepackage[all]{hypcap}  %Ensures figure hyperlinks are positioned to display entire figure
\hypersetup{
    colorlinks=true,		%false: black links; true: colored links
    linkcolor=red,		%color of internal links
    citecolor=blue,		%color of links to bibliography
    filecolor=magenta,	%color of file links
    urlcolor=blue	%color of external links
}
\newcommand{\IR}{\text{\sc \fontfamily{ptm}\selectfont ir}}
\newcommand{\UV}{\text{\sc \fontfamily{ptm}\selectfont uv}}
\newcommand{\WCSM}{\text{\sc \fontfamily{ptm}\selectfont wcsm}}
\newcommand{\LamW}{\Lambda_\text{\sc \fontfamily{ptm}\selectfont w}}
\newcommand{\alphaW}{\alpha_\text{\sc \fontfamily{ptm}\selectfont w}}
\newcommand{\tabref}[1]{Table~\ref{#1}}
\newcommand{\secref}[1]{Section~\ref{#1}}
\newcommand{\B}{\mathsf{B}}
\renewcommand{\L}{\mathsf{L}}

\begin{document}
\title{A phase of confined electroweak force in the early Universe}
\author{Joshua Berger}%\email{josh.berger@pitt.edu}
\affiliation{Department of Physics and Astronomy, University of Pittsburgh, Pittsburgh, Pennsylvania 15260, USA}
\author{Andrew J. Long}%\email{andrewjlong@rice.edu}
\affiliation{Leinweber Center for Theoretical Physics, University of Michigan, Ann Arbor, Michigan 48109, USA}
\affiliation{Department of Physics and Astronomy, Rice University, Houston, Texas 77005, USA}
\author{Jessica Turner}%\email{jturner@fnal.gov}
\affiliation{Fermi National Accelerator Laboratory,
P.O. Box 500, Batavia, Illinois 60510, USA}

\preprint{FERMILAB-PUB-19-269-T}
\preprint{PITT-PACC-1903}

\date{\today}

\begin{abstract}
We consider a modified cosmological history in which the presence of beyond-the-Standard-Model physics causes the weak gauge sector, $\SU{2}_L$, to confine before it is Higgsed.
Under the assumption of chiral symmetry breaking, quark and lepton weak doublets form condensates that break the global symmetries of the Standard Model, including baryon and lepton number, down to a $\mathrm{U}(1)$ subgroup under which only the weak singlet fermions and Higgs transform.  The weakly-coupled gauge group $\mathrm{SU}(3)_c \times \mathrm{U}(1)_Y$ is also broken to an $\mathrm{SU}(2)_c \times \mathrm{U}(1)_Q$ gauge group.  The light states include (pseudo-)Goldstone bosons of the global symmetry breaking, mostly-elementary fermions primarily composed of the weak singlet quarks and leptons, and the gauge bosons of the weakly-coupled gauge group.  We discuss possible signatures from early Universe cosmology including gravitational wave radiation, topological defects, and baryogenesis.
\end{abstract}
\maketitle

%==================
% Introduction
%==================
\section{Introduction}\label{sec:intro}

%=========
Quarks and leptons can interact via the weak force, mediated by the $W$- and $Z$-bosons.  The Standard Model of elementary particles (SM) describes the electroweak force by a quantum field theory that is constructed from the gauge group $\SU{2}_L \times \U{1}_Y$ and is in the Higgs phase~\cite{Glashow:1961tr,Weinberg:1967tq,Salam:1968rm}.  Higgsing screens weak isospin at the weak scale and leaves only weakly-coupled electromagnetic long-range interactions.  The theory is {\it weakly coupled in the infrared}, which agrees extremely well with measured quark and lepton interactions.  By contrast the strong force, which mediates interactions among the quarks, is {\it strongly coupled in the infrared}, leading to the the confinement of the constituent quarks into composite states, namely mesons and baryons.  In this article, we study the weak-confined Standard Model (WCSM) in which the $\SU{2}_L$ component of the electroweak force is strongly coupled and weak isospin is confined in composite particles.  

Confinement of $\SU{2}_L$ was studied in the pioneering work of Abbott, Farhi, and others in the 1980's~\cite{Abbott:1981re,Abbott:1981yg,Claudson:1986ch,tHooft:1998ifg} (also \cite{Calmet:2000th,Calmet:2001rp,Calmet:2001yd}) as an alternative to the SM.  Assuming that confinement occurs without chiral symmetry breaking, they argued that the known ``elementary particles'' are actually composite particles and that the predicted low-energy fermion spectrum agrees well with the measured quark and lepton masses.  
However, this scenario is now ruled out by precision measurements of the electroweak sector~\cite{DEramo:2009eqs}, and it would appear that the weak force is not confining in our present day universe.  

Nevertheless, when we study particle physics in a cosmological context, the system can exist in different phases at different times, corresponding to different temperatures of the primordial plasma.  We explore a scenario in which the $\SU{2}_L$ weak force was strongly coupled {\it in the early Universe}, when the plasma temperature was $T \gg 100 \GeV$, but became weakly coupled and Higgsed well before the epoch of nucleosynthesis, $T \sim \mathrm{MeV}$. Although we are unable to access the weak-confined phase in the laboratory today, {\it e.g.}\ at collider experiments, the Universe may contain relics from this period in its cosmic history, such as gravitational waves and the baryon asymmetry of the Universe.  

Our work shares some common elements with several previous studies of SM exotic phases, their associated phase transitions, and their imprints on cosmology.  As we have already mentioned above, $\SU{2}_L$ confinement was studied in Refs.~\cite{Abbott:1981re,Abbott:1981yg,Claudson:1986ch,tHooft:1998ifg,Calmet:2000th}, and we present a detailed comparison of our work with theirs in \sref{sec:AbbottFarhi}.  The authors of Refs.~\cite{Samuel:1999am,Quigg:2009xr} studied a variation of the SM in which the Higgs field is absent, and the electroweak force is Higgsed by the quark chiral condensates, and Ref.~\cite{Bai:2018vik} points out a dark matter candidate in a related model.  Delaying the electroweak phase transition (EWPT) to coincide with the chiral phase transition of QCD was studied as a means of generating the baryon asymmetry of the Universe \cite{Kuzmin:1992up,Servant:2014bla,vonHarling:2017yew} (see also \rref{Ellis:2019flb}) and a stochastic background of gravitational waves~\cite{Iso:2017uuu}.  Likewise, an early period of QCD confinement, which triggers a first order EWPT, was proposed as a possible mechanism of baryogenesis~\cite{Ipek:2018lhm}.  Moreover, early QCD phase transitions have been studied in order to open the axion parameter space~\cite{Dvali:1995ce,Choi:1996fs}.  

We will see that $\SU{2}_L$ confinement in the WCSM causes baryon number and lepton number to be spontaneously broken, which is signaled by the formation of lepton-quark and quark-quark condensates, $\langle l q \rangle$ and $\langle q q \rangle$.  
The violation of baryon number is a necessary condition to explain the cosmological excess of matter over antimatter, and therefore one may expect that the WCSM can naturally accommodate baryogenesis.  
However, when we discuss the cosmological implications of weak confinement in \sref{sec:cosmoimplications}, we will see that the viability of baryogenesis remains unclear.  

This article is structured as follows.  In \sref{sec:weakconfinement} we provide an example of the new physics that could lead to weak confinement in the early universe.  In order to develop intuition for confinement in an $\SU{2}$ gauge theory, we next study three simplified models in \sref{sec:toymodels}.  Turning to the WCSM in \sref{sec:WCSM}, we investigate the symmetry breaking pattern, calculate the spectrum of composite particles, and discuss their interactions.  We discuss the possible cosmological implications in \secref{sec:cosmoimplications} and highlight directions for future work.  We contrast the WCSM with Abbot and Farhi's earlier work on $\SU{2}_L$ confinement in \sref{sec:AbbottFarhi} and conclude in \sref{sec:Summary}.  

%==================
% Confinement of the Weak Force
%==================
\section{Confinement of the Weak Force}\label{sec:weakconfinement}

%=========
What new physics might allow the $\SU{2}_L$ weak force to confine in the early Universe?  
In the context of QCD, early color confinement has been studied previously in Refs.~\cite{Dvali:1995ce,Choi:1996fs} (see also \rref{Ipek:2018lhm}), and we can straightforwardly adapt that mechanism for our purposes.  
The essential idea is to link the strength of the weak force with the expectation value of a scalar field that experiences a phase transition in the early Universe.  

%=========
Let $\hat{\varphi}(x)$ be a real scalar field, which we call the modulus field, and suppose that it interacts with the weak force through a dimension-five operator.  
The relevant Lagrangian is 
\begin{align}\label{eq:L_modulus}
	\Lscr = - \frac{1}{2} \left( \frac{1}{g^2} - \frac{\hat{\varphi}}{M} \right) \mathrm{Tr}\bigl[ W_{\mu\nu} W^{\mu\nu} \bigr] - V(\hat{\varphi})
	\com 
\end{align}
where $g$ is the $\SU{2}_L$ gauge coupling, $M$ is an energy scale parameter, $W$ is the $\SU{2}_L$ field strength tensor, and $V$ is the scalar potential of $\hat{\varphi}$.  
In the limit $M \to \infty$ the modulus field decouples from the Standard Model.  
We will consider $M > \mathrm{TeV}$ in order to avoid disrupting the remarkable agreement between electroweak-precision theory and measurement.  

%=========
In general the scalar modulus may have a nonzero expectation value, $\langle\hat{ \varphi} \rangle$, and in this background, the strength of the $\SU{2}_L$ interaction is controlled by the effective coupling
\begin{align}\label{eq:geffsq}
	\frac{1}{g_\mathrm{eff}^2} = \frac{1}{g^2} - \frac{\langle\hat{ \varphi} \rangle}{M}
	\per
\end{align}
If $\langle\hat{ \varphi} \rangle \ll M / g^2$ then $g_\mathrm{eff} \approx g$, but if $\langle\hat{ \varphi} \rangle \sim M / g^2$ then the weak force is stronger, $g_\mathrm{eff} \gtrsim g$.  
The value of $\langle\hat{ \varphi} \rangle$ need not remain fixed throughout the cosmic history, and we demonstrate this point in the paragraphs below.  
Thus we will assume that $\langle\hat{ \varphi} \rangle$ was large in the early Universe, corresponding to $g_\mathrm{eff} > g$, but that it evolved to a small value, where $g_\mathrm{eff} \approx g$, at some point before the epoch of big bang nucleosynthesis.  
In this way, laboratory probes of the weak force only access $g_\mathrm{eff} \approx g$, whereas cosmological probes of the pre-nucleosynthesis era may uncover evidence for $g_\mathrm{eff} > g$ in the early Universe.  

\begin{figure}[t]
\begin{center}
\includegraphics[width=0.47\textwidth]{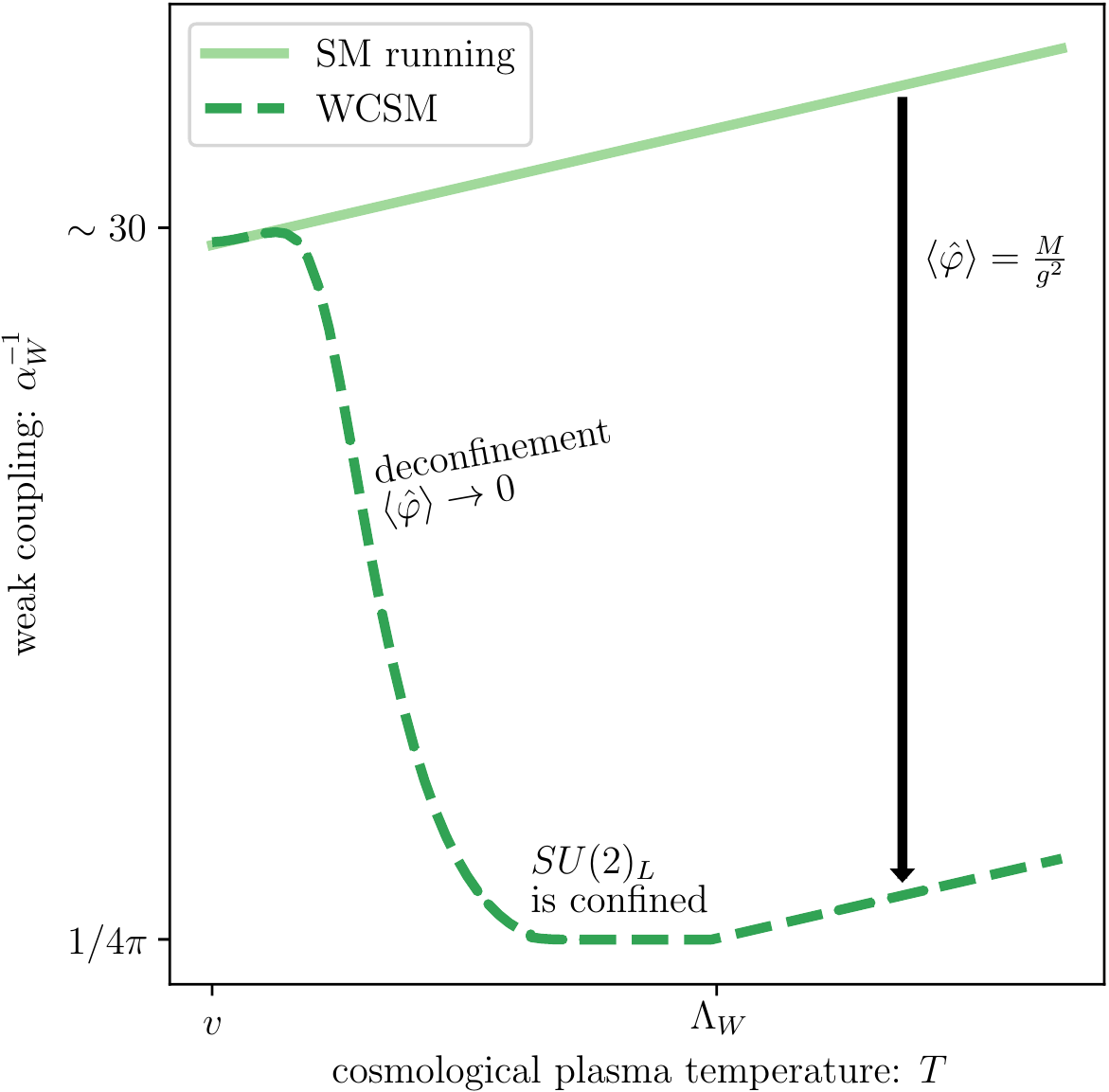} 
\caption{\label{fig:running}
The strength of the $\SU{2}_L$ weak force is parametrized by the fine structure constant $\alphaW = g_\mathrm{eff}^2 / 4\pi$, which varies with the temperature of the cosmological plasma, $T$, according to the renormalization group flow.  The solid-green line illustrates the Standard Model prediction:  the weak force grows stronger (smaller $\alphaW^{-1}$) as the plasma cools (smaller $T$), but it remains weak ($\alphaW^{-1} \gg 1/4\pi$) when the theory is Higgsed at $T \sim v \sim 100 \GeV$.  Instead, we are interested in the weak-confined Standard Model, corresponding to the dashed-green line:  the nonzero modulus field makes the weak force stronger (smaller $\alphaW^{-1}$) at early times (large $T$) such that $\alphaW$ reaches $\sim 4\pi$ at $T \sim \LamW$, and the theory enters the $\SU{2}_L$-confined phase.  
}
\end{center}
\end{figure}

%=========
As  discussed, the effective strength of the weak interaction, $g_\mathrm{eff}$, varies with time through the scalar field modulus, $\langle\hat{ \varphi} \rangle$, but it also varies through the cosmological plasma temperature, $T$.  
At temperatures $T > 100 \GeV$, the hot Standard Model plasma contains $\SU{2}_L$-charged quarks and leptons, which interact via the weak force.  
As the Universe expands and the cosmological plasma cools, these scatterers carry less kinetic energy, and they probe the weak force at larger-and-larger length scales.  
This scale dependence of the weak force is described by the equations of renormalization group flow, and its effects are captured by treating $g$ as a running coupling that is tiny at small length scales in the high-energy ultraviolet (UV) regime and grows bigger at large length scales in the low-energy infrared (IR) regime.  
In the Standard Model the weak force was Higgsed at the electroweak phase transition when the plasma temperature was $T \sim v \sim 100 \GeV$, corresponding to $g^2/4\pi \ll 1$.  
However, if $\langle\hat{\varphi} \rangle \sim M/g^2$ at early times, then the weak coupling may run to a non-perturbatively large value, $g_\mathrm{eff}^2 / 4\pi = \mathcal{O}(4\pi)$, while the Higgs field's expectation value remains zero.  
In that case the Standard Model enters the confined-$\SU{2}_L$ phase, which we call the WCSM, and we study the properties of this phase in the remainder of this article.  
The proposed modification to the gauge coupling's running is schematically illustrated in \fref{fig:running}.

%=========

%=========
One can imagine several different explanations for the nontrivial dynamics of $\hat{\varphi}$.  
On the one hand, the potential $V(\hat{\varphi})$ may have a local minimum at $\langle\hat{ \varphi} \rangle \neq 0$, which traps the field until it is either released, for instance by the changing temperature of the cosmological plasma, or it escapes through a first order phase transition.  
Alternatively, the potential $V(\hat{\varphi})$ may simply be very shallow such that the field $\hat{\varphi}$ is ``frozen by Hubble drag'' at its nonzero value until a time such that the Hubble parameter, $H$, decreases below its effective mass,  {\it i.e.}, $H < m_{\hat{\varphi}} \sim |V^{\prime\prime}|^{1/2}$.\footnote{It is worth remarking that the modulus field's effective potential is a sum of $V(\hat{\varphi})$ and a correction induced by confinement, which is $\Delta V \sim \LamW^4$.  The confinement scale can be written as $\LamW^4 \sim \Lambda_\UV^4 \, e^{-8\pi^2 / g_\mathrm{eff}^2}$ where $\Lambda_\UV$ is a high-energy input scale and $g_\mathrm{eff}$ depends on $\hat{\varphi}$ through \eref{eq:geffsq}.  The corresponding mass correction, $\Delta m_{\hat{\varphi}} \sim 4 \pi^2 \LamW^2 / M$, may be larger than $H \sim T^2 / \Mpl$ at the time of confinement when $T \sim \LamW$, which would cause $\hat{\varphi}$ to be released from its Hubble drag before the system spends much time in the confined phase.  This outcome can be avoided by taking $M > \Mpl$ or possibly by tuning $V(\hat{\varphi})$ against $\Delta V$.  }   
If the Universe is radiation-dominated at this time, then $T \sim \sqrt{m_{\hat{\varphi}} \Mpl}$ where $\Mpl$ is the reduced Planck mass.  
For instance if $m_{\hat{\varphi}} \sim 10^{-5} \eV$ then $T \sim 100 \GeV$.  
Once $\hat{\varphi}$ is released, it can roll toward the minimum of its potential, located at $\langle\hat{\varphi} \rangle = 0$, where it oscillates and decays.  

%=========
In the remainder of this article, we do not assume any particular model for inducing confinement of the weak force. 
Instead we assume that the $\SU{2}_L$ weak interaction becomes non-perturbatively large at a scale $\LamW \gtrsim \mathrm{TeV}$ in the early Universe at a time before the electroweak interaction is Higgsed and before color is confined.  

%==================
% Building up the to the WCSM
%==================
\section{Building up to the WCSM}
\label{sec:toymodels}

%=========
Our goal is to study the weak-confined Standard Model, which has the same particle content as the SM, but in which the $\SU{2}_L$ weak force is confined.  
As weak confinement leads to a dramatic departure from the usual SM dynamics, we use this section to put together the pieces of the $\SU{2}_L$-confined SM bit-by-bit.  
First, we consider an $\SU{2}$ gauge theory with doublet fermions followed by the addition of a doublet scalar, and we finally consider singlet fermions.  

%-----------------------
% Model 1
%-----------------------
\subsection{Model 1: Include only $\SU{2}_L$-doublet fermions}\label{sec:toymodel1}

%=========
We consider a gauge theory based on the $\SU{2}$ symmetry group, which we denote by $\SU{2}_L$, anticipating the connection with SM weak isospin.  
The force carriers are represented by the vector field $W_\mu(x)$, which transforms in the adjoint of $\SU{2}_L$, and the matter is represented by $2N_f$ flavors\footnote{An even number of fermions ($2N_f$) is required to avoid an anomaly in the $\SU{2}_L$ gauge group.  We assume $N_f > 1$, but see \rref{Francis:2018xjd} for an analysis of $N_f = 1$. } of left-chiral Weyl fermion fields, denoted by\footnote{Here and throughout the text we suppress the spinor and $\SU{2}_L$ gauge indices when appropriate, showing instead only the flavor indices.} $\psi_i(x)$ for $i = 1, 2, \cdots, 2 N_f$, which separately transform in the fundamental representation of $\SU{2}_L$.  
The fermions are assumed to have no mass terms, and the Lagrangian is simply 
\begin{align}\label{eq:L1_UV}
	\Lscr_1 \bigr|_\UV = 
	\sum_{i=1}^{2N_f} \psi_i^\dagger i \bar{\sigma}^\mu D_\mu \psi_i 
	- \frac{1}{2g^2} \, \mathrm{Tr}\bigl[ W_{\mu\nu} W^{\mu\nu} \bigr]
	\ , 
\end{align}
where $D_\mu \psi_i = \partial_\mu \psi_i - i W_\mu \psi_i$ is the covariant derivative of $\psi_i$, $g$ is the running $\SU{2}_L$ gauge coupling at the UV scale, and $W_{\mu\nu}$ is the $\SU{2}_L$ field strength tensor.\footnote{In general the Lagrangian will also contain a topological term, $\Lscr_\theta = - (\theta / 16\pi^2) \, \mathrm{Tr}[ W_{\mu\nu} \tilde{W}^{\mu\nu} ]$ where $\theta$ is the vacuum angle.  However, as we will discuss momentarily, this theory has an anomaly in axial $\psi$-number and no other explicit breaking of this symmetry, which makes it possible to perform a field redefinition and remove the topological term from the Lagrangian, implying that $\theta$ cannot affect any observable.  Analogously the weak vacuum angle can be removed from the SM Lagrangian through the baryon- or lepton-number anomalies~\cite{Perez:2014fja}.  Therefore, without loss of generality, we will work in the basis where the topological term is absent.}
In \secref{sec:WCSM} we will take $2N_f = 12$ to count the SM's nine quark doublets and three lepton doublets, but here we keep the analysis more general.

%=========
The theory of \eref{eq:L1_UV} has a $\U{2N_f}_\psi$ flavor symmetry under which the $\psi_i$ transform in the fundamental representation and $W_\mu$ transforms as a singlet.  
Using the isomorphism $\U{2N_f}_\psi = \SU{2N_f}_\psi \times \U{1}_\psi$, we can assign $\psi$ a charge $q_\psi$ under the $\U{1}_\psi$ group whose conserved charge is axial-$\psi$-number.  
The $\SU{2}_L$ gauge interactions induce an anomaly in $\U{1}_\psi$~\cite{Hooft:1976up}, breaking it to the $\mathbb{Z}_{2N_f}$ subgroup.  
The transformation properties of the various fields are summarized in \tabref{tab:charge_assignments1}.  

%=========
It is illustrative to draw a contrast with the flavor symmetry of QCD~\cite{Scherer:2002tk}.  
In the limit that the up, down, and strange quarks and antiquarks are massless, we would have $N_f = 3$ flavors of left-chiral Weyl fermions transforming in the fundamental of $\SU{3}_c$ color, ${\bm 3}$, and $N_f = 3$ additional flavors (of left-chiral Weyl fermions) transforming in the anti-fundamental, $\bar{\bm 3}$.  
The flavor symmetry is $\U{N_f} \times \U{N_f} = \SU{N_f}_V \times \U{1}_V \times \SU{N_f}_A \times \U{1}_A$ where the conserved charge associated to $\U{1}_V$ is baryon number and the anomalous charge associated with $\U{1}_A$ is axial-baryon number.  
The situation in the theory with gauged $\SU{2}_L$ is different because the fundamental representation is pseudo-real, {\it i.e.} ${\bm 2} = \bar{\bm 2}$, which allows the $2N_f$ fields to be collected into a single flavor multiplet with the larger $\U{2N_f}$ flavor symmetry instead.  

%=========
The $\SU{2}_L$ force in \eref{eq:L1_UV} is not Higgsed, as in the SM, but rather may become strongly coupled in the IR.  
To analytically understand the evolution towards strong coupling, one can calculate the gauge coupling's beta function $\beta_g$, and a value $\beta_g < 0$ implies that $g$ grows in the IR.  
Calculating $\beta_g$ at one-loop order in perturbation theory gives $\beta_g = (b / 16\pi^2) g^3$ with $b = -22/3 + 2N_f / 3$~\cite{Gross:1973id}.  
A theory with $N_f < 11$ has $\beta_g < 0$, and this perturbative analysis suggests that such a theory is strongly coupled in the IR; that is, $g^2 / 4\pi$ grows as large as $4\pi$ as the energy scale decreases toward $\LamW$.  
This perturbative analysis is confirmed by numerical lattice studies for small $N_f$ while $N_f = 6$ is marginal and near the conformal window~\cite{Karavirta:2011zg,Hayakawa:2013maa,Amato:2015dqp,Leino:2017hgm,Leino:2018qvq}.  

%=========
The strong coupling is assumed to imply that $\SU{2}_L$ charge is confined, in analogy with QCD's color confinement.\footnote{If the theory instead has a strongly-coupled, IR-stable fixed point, as suggested by \rref{Leino:2017hgm}, then there is no charge confinement, and a different analysis is required to determine the properties and interactions of the low-energy degrees of freedom.  }  
Consequently the degrees of freedom in the low-energy confined phase are composite states, analogs of the mesons and baryons of QCD.  
Specifically, the spectrum consists of several massless Goldstone bosons $\Pi^a$, associated with the assumed $\SU{2N_f}_\psi$ global flavor symmetry, a single massive pseudo-Goldstone boson $\eta^\prime$, associated with the anomalous $\U{1}_\psi$ global flavor symmetry, and a tower of heavy resonances with masses around the confinement scale, $m_\mathrm{res} \sim \LamW$.  
These heavy resonances include 
meson-like scalars, vectors, and glueball-like bound states of the $W_\mu$ boson.  
As these resonances are expected to be heavy and unstable, we can assume that they decay shortly after $\SU{2}_L$ is confined, and we focus on the properties of the light states.  

%=========
Collectively the lowest-lying, meson-like states $\Pi^a$ and $\eta^\prime$ are encoded in the scalar fields $\Sigma_{ij}(x)$, which have the same symmetry properties as the fermion bilinears:
\begin{align}\label{eq:Sigma_to_psi}
	\Sigma_{ij} \sim \psi_i \varepsilon \psi_j
	\per
\end{align}
Here the totally-antisymmetric tensor, $\varepsilon$, with $\varepsilon_{12} = 1$, contracts the (suppressed) $\SU{2}_L$ indices such that $\Sigma_{ij}$ is an $\SU{2}_L$ singlet.  
Under the flavor symmetries, $\Sigma$ transforms as an antisymmetric rank-2 tensor of $\SU{2N_f}_\psi$ with charge $2 q_\psi$ under $\U{1}_\psi$.  
We assume that $\Sigma$ acquires a nonzero vacuum expectation value, and the global $\SU{2N_f}_\psi \times \U{1}_\psi$ symmetry is spontaneously broken.  

%=========
 The symmetry breaking pattern cannot be determined through a perturbative calculation, since after all, symmetry breaking is a consequence of non-perturbative charge confinement.  
However, this simple model has been studied using non-perturbative, numerical lattice techniques~\cite{Lewis:2011zb,Arthur:2016dir} for $N_f = 2$ (see also Refs.~\cite{Karavirta:2011zg,Hayakawa:2013maa,Amato:2015dqp,Leino:2017hgm,Leino:2018qvq}).  
Those studies conclude that $\SU{2N_f}_\psi$ is broken to $\Sp{2N_f}_\psi$, which is the naive expectation from chiral symmetry breaking in analogy with QCD~\cite{Preskill:1980mz,Kosower:1984aw}.  
Therefore, we will adopt the results of the lattice studies, and assume that the unbroken global flavor symmetry is the $\Sp{2N_f}_\psi$ group.  
This pattern of symmetry breaking is obtained if the antisymmetric field $\Sigma_{ij}(x)$ acquires a vacuum expectation value $\langle \Sigma_{ij} \rangle = (\Sigma_0)_{ij}$ that satisfies 
\begin{align}\label{eq:Sigma_vev}
	\Sigma_0^\dagger \Sigma_0 = \Sigma_0 \Sigma_0^\dagger = \bbone 
	\quad \Leftrightarrow \quad 
	\SU{2N_f}_\psi \, / \, \Sp{2N_f}_\psi 
	\per
\end{align}
Without loss of generality, we work in the basis where $\Sigma_0$ is a block diagonal matrix with $N_f$ blocks of $\varepsilon$ along the diagonal.  

%=========
Due to the spontaneous symmetry breaking in \eref{eq:Sigma_vev}, Goldstone's theorem implies that there will be massless Goldstone bosons in the spectrum of low-lying, meson-like states.  
Note that $\SU{2N_f}_\psi$ has $4N_f^2 - 1$ generators, which we denote by $T_{ij}^a$ and $X_{ij}^a$ such that $T_{ij}^a$ for $a = 1, 2, \cdots, 2N_f^2 + N_f$ are the generators of the unbroken subgroup $\Sp{2N_f}_\psi$ and $X_{ij}^a$ for $a = 1, 2, \cdots, 2N_f^2 - N_f - 1$ are the generators of the broken symmetries. 
The $\U{1}_\psi$ factor is also broken by \eref{eq:Sigma_vev}.  
Thus we expect the spectrum to contain $2N_f^2 - N_f - 1$ massless Goldstone bosons, denoted by $\Pi^a$, and one massive pseudo-Goldstone boson, denoted by $\eta^\prime$; the notation is chosen to draw an analogy with the pions and $\eta^\prime$ meson of QCD.  
Since $\U{1}_\psi$ is anomalous, the $\eta^\prime$ mass is lifted by $\SU{2}_L$-instanton effects (see below).   

%=========
In the confined phase, the low-energy physics can be described with a non-linear sigma model, 
\begin{align}\label{eq:L1_IR}
	\Lscr_1 \bigr|_\IR & = 
	\frac{f^2}{4} \mathrm{Tr}\bigl[ \partial_\mu \Sigma^\dagger \partial^\mu \Sigma \bigr] 
	+ \kappa \LamW^2 f^2 {\rm Re}[{\rm det} \, \Sigma]
	\com
\end{align}
where $f$ is the decay constant, $\kappa$ is an $\mathcal{O}(1)$ dimensionless number, and $\LamW \sim 4 \pi f$ is the confinement scale.  
As we have discussed above, the antisymmetric complex scalar field $\Sigma_{ij}(x)$ parametrizes the massless Goldstone boson fields $\Pi^a(x)$ and the pseudo-Goldstone boson field $\eta^\prime(x)$.  
We make this connection explicit by writing 
\begin{align}\label{eq:Sigma_def}
	\Sigma = 
	\mathrm{exp}\bigl[i \eta^\prime / \sqrt{N_f} f \bigr]
	\ \mathrm{exp}\bigl[ \sum_{a} 2 i X^a \Pi^a / f \bigr] \ \Sigma_{0}
	\com
\end{align}
where $X_{ij}^a$ are the $2N_f^2 - N_f - 1$ broken generators of $\SU{2N_f}$, and the vacuum expectation value $\Sigma_0$ satisfies \eref{eq:Sigma_vev}.  
The factors of $2$ and $\sqrt{N_f}$ are included to ensure canonical normalization of 
 kinetic terms  and the generators are normalized such that $\mathrm{Tr}[X^a X^b] = \delta^{ab}/2$.  

%=========
The last term of \eref{eq:L1_IR} arises from instanton effects via the chiral anomaly of $\U{1}_\psi$.  
The anomaly explicitly breaks the $\U{1}_\psi$ symmetry to its $\mathbb{Z}_{2N_f}$ subgroup and thereby lifts the mass of the $\eta^\prime$ through instanton effects, in analogy with the $\U{1}_A$ problem of QCD~\cite{Hooft:1986nc}.  
One calculates the $\eta^\prime$ mass by substituting \eref{eq:Sigma_def} into \eref{eq:L1_IR} and expanding to quadratic order; doing so gives $m_{\eta^\prime}^2 = 4 \kappa N_f \LamW^2$.  

%-----------------------
% Model 2
%-----------------------
\subsection{Model 2: Add an $\SU{2}_L$-doublet scalar}\label{sec:toymodel2}

%=========
Let us next consider an extension of the first model that is a closer approximation to the Standard Model particle content, by introducing an analog of the Higgs doublet field.  
Consider the complex scalar field $\phi(x)$ that transforms under $\SU{2}_L$ in the fundamental representation. The Lagrangian of the UV degrees of freedom can be written as (see also \rref{Hambye:2009fg})
\begin{align}\label{eq:L2_UV}
	\Lscr_2 \bigr|_\UV = \Lscr_1 \bigr|_\UV + \bigl| D_\mu \phi \bigr|^2 - U( \phi^\ast \phi )
	\com
\end{align}
where $D_\mu \phi = \partial_\mu \phi - i W_\mu \phi$ is the covariant derivative, and $U$ is a potential for the scalar $\phi$.  
We require the Lagrangian to be invariant under $\SU{2}_L$, which forces $U$ to be a function of $\phi^\ast \phi$ only, since $\phi \varepsilon \phi$ is identically zero, and forbids renormalizable interactions between $\phi$ and the $\psi_i$.  

%=========
The complex $\SU{2}_L$-doublet field $\phi$ consists of $4$ real degrees of freedom, and in fact \eref{eq:L2_UV} respects an $\SO{4}$ symmetry. It is useful to apply the isomorphism $\SO{4} = \SU{2}_L \times \SU{2}_\phi$ where the first factor is identified as the gauged subgroup, and the second factor is a new global flavor symmetry which is the analog of custodial $\SU{2}$ in the Standard Model.  
Thus the global flavor symmetry is enlarged to $\U{2N_f}_\psi \times \SU{2}_\phi$.  
It is useful to recall that an $\SU{2}_L$-doublet field $\phi(x)$ can also be represented as $\varphi_a(x)$ with~\cite{Sikivie:1980hm}
\begin{align}\label{eq:varphi_def}
	\varphi_1 = \varepsilon \phi
	\qquad \text{and} \qquad 
	\varphi_2 = - \phi^\ast 
	\com
\end{align}
where $\varphi$ transforms as ${\bm 2}$ under each factor of $\SU{2}_L \times \SU{2}_\phi$.  
Using this equivalent representation of the $\SU{2}_L$-doublet scalar, we can write the Lagrangian as 
\begin{align}\label{eq:L2_UV_alt}
	\Lscr_2 \bigr|_\UV = \Lscr_1 \bigr|_\UV + \frac{1}{2} \sum_{a=1,2} \bigl| D_\mu \varphi_a \bigr|^2 - U( |\varphi|^2 / 2 )
	\com
\end{align}
where $D_\mu \varphi_a = \partial_\mu \varphi_a + i W_\mu \varphi_a$, and we have used the shorthand notation $|\varphi|^2 \equiv \varphi_1^\dagger \varphi_1 + \varphi_2^\dagger \varphi_2$.  

%=========
From the low energy perspective, we have extended the spectrum by additional low-lying states, corresponding to a new real scalar field $\Phi(x)$ and a set of new Weyl fermions $\Psi_{ia}(x)$ for $i = 1, 2, \cdots, 2N_f$ and $a = 1,2$, which have the same symmetry properties as the following bilinears 
\begin{subequations}\label{eq:Phi_Psi}
\begin{align}
	\Phi & \sim \phi^\ast \phi \\ 
	\Psi_{ia} & \sim \psi_i \varphi_a 
	\com
\end{align}
\end{subequations}
which are all singlets under $\SU{2}_L$.  
The composite scalar, $\Phi$, is a total singlet, while the composite fermions, $\Psi_{ia}$, transform as fundamentals under the $\SU{2N_f}_\psi$ flavor symmetry, carry charge $q_\psi$ under the $\U{1}_\psi$ symmetry, and transform as a ${\bm 2}$ under the $\SU{2}_\phi$ custodial symmetry as  summarized in \tabref{tab:charge_assignments1}.  
Note that $\Phi$ is a singlet under the custodial $\SU{2}$, and there does not exist an analogous triplet state; only the custodial anti-symmetric combination is allowed.  

%=========
The low energy physics is described by 
\begin{align}\label{eq:L2_IR}
	\Lscr_2 \bigr|_\IR & = 
	\Lscr_1 \bigr|_\IR 
	+ 
	\sum_{i=1}^{2N_f} \sum_{a=1,2} \Psi_{ia}^\dagger i \bar{\sigma}^\mu \partial_\mu \Psi_{ia} 
	\\ & \quad 
	+ \frac{1}{2} \partial^\mu \Phi \partial_\mu \Phi 
	- \LamW^2 f^2 \, V(\Phi / f)
	\nn & \quad 
	- \sum_{i,j=1}^{2N_f} \sum_{a,b=1,2} \bigl[ y_0 \LamW \varepsilon^{ab} \Psi_{ia} \Sigma_{ij}^\dagger \Psi_{jb} + \hc \bigr] 
	\nn & \quad 
	- \sum_{i,j=1}^{2N_f} \sum_{a,b=1,2} \bigl[ ( y_0^\prime \LamW / f) \varepsilon^{ab} \Psi_{ia} \Sigma_{ij}^\dagger \Psi_{jb} \Phi + \hc \bigr] 
	\com \nonumber
\end{align}
where $f$ is the decay constant, $\LamW \sim 4 \pi f$ is the confinement scale, $V$ is a potential for the composite scalar $\Phi$, and $y_0$ and $y_0^\prime$ are Yukawa couplings.  
The $4\pi$ counting is performed using na\"ive dimensional analysis~\cite{Weinberg:1978kz,Manohar:1983md,Cohen:1997rt},
which implies that $y_0$, $y_0^\prime$, and the dimensionless coefficients in $V$ are $\mathcal{O}(1)$ numbers.  
We assume that the potential $U$ in \eref{eq:L2_UV} is a small perturbation on $V$, such that the scalar's mass is of order $m_\Phi \sim \LamW$.
The potential $V$ may also induce a nonzero vacuum expectation value for $\Phi$, which is expected to be $\langle \Phi \rangle \sim f$, but since $\Phi$ is a singlet, this vacuum expectation value does not signal any spontaneous symmetry breaking.  
The $y_0$ term, which is a Yukawa interaction, respects the $\U{2N_f}_\psi \times \SU{2}_\phi$ flavor symmetry, and when $\Sigma_{ij}$ acquires a nonzero vacuum expectation value \pref{eq:Sigma_vev}, it induces a mass term for the $\Psi_{ia}$.  
Together $\Psi_{ia}$ combine in pairs to form $2N_f$ Dirac fermions, which are all degenerate and have a mass $m_\Psi = y_0 \LamW$.  

%-----------------------
% Model 3
%-----------------------
\subsection{Model 3: Add $\SU{2}_L$-singlet fermions}\label{sec:toymodel3}

%=========
As a final iteration, we introduce a set of $\SU{2}_L$-singlet, left-chiral Weyl fermion fields, denoted as $\xi_i(x)$ for $i = 1, 2, \cdots, N_\xi$.  
Note that $N_\xi$ may be either even or odd, since these fields do not contribute to the $\SU{2}_L$ anomaly.  
As we build up to the SM, these fields will correspond to the $N_\xi = 21$ $\SU{2}_L$-singlet quark and lepton fields.  

%=========
We require the theory to respect an extended global symmetry group, $\U{2N_f}_\psi \times \SU{2}_\phi \times \U{N_\xi}_\xi$ where the $\xi_i$ transform in the fundamental representation of the $\U{N_\xi}_\xi$ and the $\psi_i$, $W_\mu$, and $\phi$ are singlets; see \tabref{tab:charge_assignments1}.  
This symmetry forbids masses such as $\mu_{ij} \xi_i \xi_j$ and interactions such as $\lambda_{ij} \phi \psi_i \xi_j$.  
Therefore the Lagrangian for the high-energy degrees of freedom is simply 
\begin{align}\label{eq:L3_UV}
	\Lscr_3 \bigr|_\UV & = \Lscr_2 \bigr|_\UV 
	+ \sum_{i=1}^{N_\xi} \xi_i^\dagger i \bar{\sigma}^\mu \partial_\mu \xi_i 
	\per
\end{align}
Moreover, these $\SU{2}_L$-singlet fermions do not feel the strong $\SU{2}_L$ force and  therefore do not participate in confinement.  
Thus the Lagrangian for the low-energy degrees of freedom is also simply 
\begin{align}\label{eq:L3_IR}
	\Lscr_3 \bigr|_\IR & = \Lscr_2 \bigr|_\IR 
	+ \sum_{i=1}^{N_\xi} \xi_i^\dagger i \bar{\sigma}^\mu \partial_\mu \xi_i 
	\per
\end{align}
These singlet fermions are entirely inert and decoupled. In the next section we find that breaking the large global flavor symmetry allows for Yukawa interactions, such as $\Delta \Lscr \sim \Psi \xi$ and $\Psi \xi \Phi$.  These interactions will have interesting consequences for the spectrum in the IR.

%=========
\begin{table}[t]
\begin{tabular}{|c||c||c|c|c|c|c|}
\hline
& \multicolumn{2}{|c|}{$\SO{4}$} & \multicolumn{2}{|c|}{$\U{2N_f}_\psi$} & \multicolumn{2}{|c|}{$\U{N_\xi}_\xi$} \\
\cline{2-7}
& $\SU{2}_L$ & $\SU{2}_\phi$ & \multicolumn{1}{|c|}{$\SU{2N_f}_\psi$} & $\U{1}_\psi$ & $\SU{N_\xi}_\xi$ &$ \U{1}_\xi$ \\ \hline 
%---
$\psi$ & {$\bm 2$} & {$\bm 1$} & \multicolumn{1}{|c|}{$\bm{2N_f}$} & $q_\psi$ & {$\bm 1$} & $$0$$ \\ 
$W_\mu$ & {$\bm 3$} & {$\bm 1$} & \multicolumn{1}{|c|}{$\bm 1$} & $0$ & {$\bm 1$} & $0$ \\ 
$\phi / \varphi$ & {$\bm 2$} & {$\bm 2$} & \multicolumn{1}{|c|}{$\bm 1$} & $0$ & {$\bm 1$} & $0$ \\ 
$\xi$ & {$\bm 1$} & {$\bm 1$} & \multicolumn{1}{|c|}{$\bm 1$} & $0$ & $\bm{N_\xi}$ & $q_\xi$ \\ \hline \hline 
$\Sigma$ & {$\bm 1$} & {$\bm 1$} & $\bm{2N_f^2 - N_f}$ & $2q_\psi$ & {$\bm 1$} & $0$ \\ 
$\Phi$ & {$\bm 1$} & {$\bm 1$} & {$\bm 1$} & $0$ & {$\bm 1$} & $0$ \\ 
$\Psi$ & {$\bm 1$} & {$\bm 2$} & $\bm{2N_f}$ & $q_\psi$ & {$\bm 1$} & $0$ \\ 
$\xi$ & {$\bm 1$} & {$\bm 1$} & {$\bm 1$} & $0$ & $\bm{N_\xi}$ & $q_\xi$ \\ \hline
\end{tabular}
\caption{\label{tab:charge_assignments1}This table summarizes the charge assignments for the sequence of models in \secref{sec:toymodels}.  Note that the $\bm{N}$- and $(\bm{N^2-N})$-dimensional representations of $\SU{N}$ are the fundamental and antisymmetric representations, respectively.  When applied to the WCSM in \secref{sec:WCSM} we take $2N_f = 12$ and $N_\xi = 21$.  
}
\end{table}

%==================
% Symmetries
%==================
\section{\label{sec:WCSM}The Weak-Confined Standard Model}

%=========
Let us now investigate weak isospin confinement in the WCSM.  
The UV particle content and symmetries of the WCSM are identical to the SM, shown in \tabref{tab:charge_assignments2}.
In particular the SM symmetry group is 
\begin{align}\label{eq:SM_sym}
	& \bigl\{ \SU{3}_c \times \SU{2}_L \times \U{1}_Y \bigr\}_\mathrm{gauge} 
	\\ & \qquad
	\times \bigl\{ \U{1}_{\Bsf + \Lsf} \bigr\}_\mathrm{anomaly}
	\times \bigl\{ \U{1}_{\Bsf/3-\Lsf_i}^3 \bigr\}_\mathrm{global} 
	\com \nonumber 
\end{align}
where the $\U{1}_{\Bsf + \Lsf}$ factor is anomalous under the $\SU{2}_L \times \U{1}_Y$ gauge interactions.
The key difference between the WCSM and the SM has to do with the strength of the $\SU{2}_L$ weak force, which is assumed to be larger in the WCSM, $g \to g_\mathrm{eff} \gg g$, such that weak isospin is confined in the IR.  
Therefore, whereas the SM predicts that $\SU{2}_L$ is Higgsed by the vacuum expectation value of $H$, the WCSM predicts that the $\SU{2}_L$ weak force is confined in the IR and the $\SU{2}_L$-doublet fields ($q_i$, $l_i$, and $H$) form composite states.  

%=========
\begin{table}[t]
\begin{tabular}{|c||ccc|cc|}
\hline
& \multicolumn{3}{c|}{\text{gauge}} & \multicolumn{2}{c|}{\text{global}} \\ 
\hline
	&$\SU{3}_c$ & $\SU{2}_L$ & $\U{1}_Y$ & $\U{1}_{\B+\L}$ & $\U{1}_{\B/3-\L_j}$ \\ \hline
	$q_i$ & {$\bm 3$} & {$\bm 2$} & $1/6$ & $1/3$ & $1/9$ \\ 
	$U_i$ & $\bar{\bm 3}$ & {$\bm 1$} & -$2/3$ & -$1/3$ & -$1/9$ \\ 
	$D_i$ & $\bar{\bm 3}$ & {$\bm 1$} & $1/3$ & -$1/3$ & -$1/9$ \\ 
	$l_i$ & {$\bm 1$} & {$\bm 2$} & -$1/2$ & $1$ & -$\delta_{ij}$ \\ 
	$E_i$ & {$\bm 1$} & {$\bm 1$} & $1$ & -$1$ & $\delta_{ij}$ \\ 
	$H$ & {$\bm 1$} & {$\bm 2$} & $1/2$ & $0$ & $0$ \\ 
	$g_\mu$ & {$\bm 8$} & {$\bm 1$} & $0$ & $0$ & $0$ \\ 
	$W_\mu$ & {$\bm 1$} & {$\bm 3$} & $0$ & $0$ & $0$ \\ 
	$B_\mu$ & {$\bm 1$} & {$\bm 1$} & $0$ & $0$ & $0$ \\	\hline
	\end{tabular} 
\caption{\label{tab:charge_assignments2}
This table shows the WCSM particle content and charge assignments.  The index $i \in \{1,2,3\}$ labels the three generations of fermions.  There are four global charges, corresponding to $\U{1}_{\B+\L}$ and $\U{1}_{\B/3-\L_j}$ for $j=1,2,3$.  All fermions are represented by left-chiral Weyl spinor fields in the two-component notation~\cite{Dreiner:2008tw}.   }
\end{table}

%=========
This section contains the main results of our paper.  
To simplify the analysis, we will begin in \sref{sec:turnoffcouplings} by setting to zero the strong coupling $g_s = 0$, the hypercharge coupling $g^\prime = 0$, and the Yukawa couplings $\lambda_\mathrm{Yuk} = 0$.  
We reintroduce these couplings in \sref{sec:turnoncouplings} where they are treated as perturbations around an enhanced symmetry point.  

%-----------------------
% Turn off strong, hypercharge, and Yukawa couplings
%-----------------------
\subsection{Turn off strong, hypercharge, and Yukawa couplings}\label{sec:turnoffcouplings}

%=========
Upon taking $g_s = g^\prime = \lambda_\mathrm{Yuk} = 0$, the symmetry group of the SM Lagrangian is enlarged from \eref{eq:SM_sym} to 
\begin{align}\label{eq:WCSM_U12}
	& \bigl\{ \SU{2}_L \bigr\}_\mathrm{gauge} \times \bigl\{ \U{1}_\psi \bigr\}_\mathrm{anomaly}
	\\ & \qquad
	\times \bigl\{ \SU{12}_\psi \times \SU{2}_\phi \times \U{21}_\xi \bigr\}_\mathrm{global} 
	\com \nonumber 
\end{align}
where the $\U{1}_\psi$ factor is anomalous under the $\SU{2}_L$ gauge interaction.  
To make these symmetries manifest in the Lagrangian, it is useful to collect the fields together into the following multiplets
\begin{align}\label{eq:psi_and_chi}
	\psi & = \{ l_{1} \, , \, q_1^R \, , \, q_1^G \, , \, q_1^B \, , \, \cdots \, , \, \cdots \} , \\ 
	\varphi & = \{ \varepsilon H \, , \, - H^\ast \}, 
	\nn 
	\xi & = \{ E_1, \, U_1^{\bar{R}} , \, U_1^{\bar{G}} , \, U_1^{\bar{B}} , \, 
%	\nn & \qquad 
	D_1^{\bar{R}} , \, D_1^{\bar{G}} , \, D_1^{\bar{B}} , \, \cdots , \, \cdots \} 
	\nonumber
	\com
\end{align}
where $\psi_i$ for $i = 1, 2, \cdots, 12$ contains the $\SU{2}_L$-doublet fermions, $\varphi_a$ for $a=1,2$ contains the $\SU{2}_L$-doublet Higgs, and $\xi_i$ for $i = 1, 2, \cdots, 21$ contains the $\SU{2}_L$-singlet fermions.  
The subscript $1$ indicates first-generation fields, the dots correspond to the second and third generations, the $\SU{2}_L$ index is suppressed, and the $\SU{3}_c$ indices are $\{ R, G, B \}$ and $\{ \bar{R}, \bar{G}, \bar{B} \}$.  
The symmetry transformation properties of these fields are shown in \tabref{tab:charge_assignments1} where $2N_f = 12$ and $N_\xi = 21$; specifically, the $\SU{2}_L$-doublet fermions transform as a $\psi_i \sim \bm{12}$ of $\U{12}_\psi$, the $\SU{2}_L$-doublet Higgs transforms as a $\varphi_a \sim \bm{2}$ of the ``custodial'' $\SU{2}_\phi$, and the $\SU{2}_L$-singlet fermions transform as a $\xi_i \sim \bm{21}$ of $\U{21}_\xi$.  

%=========
Using the multiplets in \eref{eq:psi_and_chi}, the WCSM Lagrangian (for $g_s = g^\prime = \lambda_\mathrm{Yuk} = 0$) is written compactly as 
\begin{align}\label{eq:L_SM_UV}
	\Lscr_\WCSM \bigr|_\UV & = \sum_{i=1}^{12} \psi_i^\dagger i \bar{\sigma}^\mu D_\mu \psi_i + \sum_{i=1}^{21} \xi_i^\dagger i \bar{\sigma}^\mu \partial_\mu \xi_i 
	\\ & \quad 
	- \frac{1}{2 g_{\text{eff}}^2} \mathrm{Tr} \bigl[ W_{\mu\nu} W^{\mu\nu} \bigr] 
%	+ D^\mu H^\ast D_\mu H 
	+ \sum_{a=1,2} \frac{1}{2} D^\mu \varphi_a^\dagger D_\mu \varphi_a 
	\nn & \quad 
	- \frac{1}{2} \mu^2 |\varphi|^2 - \frac{1}{4} \lambda |\varphi|^4 
	\nonumber 
	\per
\end{align}
Here we have written the $\SU{2}_L$ gauge coupling as $g_\mathrm{eff} > g$, which distinguishes the WCSM from the SM.  
Additionally, $D_\mu \psi_i = \partial_\mu \psi_i - i W_\mu \psi_i$ is the covariant derivative of the doublet fermions, $D_\mu \varphi_a = \partial_\mu \varphi_a + i W_\mu \varphi_a$ 
is the Higgs field's covariant derivative, $W_{\mu\nu}$ is the $\SU{2}_L$ field strength tensor,  $\mu^2$ is the Higgs mass parameter, $\lambda$ is the Higgs self-coupling, and $|\varphi|^2 \equiv \varphi_1^\dagger \varphi_1 + \varphi_2^\dagger \varphi_2$.  
We note that the gluon and hypercharge bosons are present but decoupled due to the assumption that their gauge couplings are zero valued.  

%=========
By neglecting the strong, hypercharge, and Yukawa interactions, we can write the Lagrangian as in \eref{eq:L_SM_UV}, which now bears a marked similarity to the models that we have already discussed in \sref{sec:toymodels} for $N_f = 6$ and $N_\xi = 21$.  
As before, we assume that the weak force confines, and the chiral symmetry is spontaneously broken.  
By applying what we learned in \sref{sec:toymodels}, we can immediately infer the effects of $\SU{2}_L$-charge confinement in the WCSM.  
The $\SU{2}_L$-doublet fermions condense in pairs to form composite scalars, $\Sigma_{ij} \sim \psi_i \varepsilon \psi_j$; the doublet Higgs condenses with the doublet fermions to form composite fermions, $\Psi_{ia} \sim \psi_i \varphi_a$; the Higgs condenses in pairs to form another composite scalar, $\Phi \sim \sum_a \varphi_a^\dagger \varphi_a \sim H^\dagger H$; and the $\SU{2}_L$-singlet fermions, $\xi_i$, naturally do not confine but (most of them) acquire masses indirectly through weak-isospin confinement as outlined in \secref{sec:turnoncouplings}.  
The symmetry transformation properties of these low-energy degrees of freedom are also shown in \tabref{tab:charge_assignments1}.  

%=========
By carrying over the results from \sref{sec:toymodels}, the Lagrangian describing the low-energy degrees of freedom of the WCSM can be written as 
\begin{align}\label{eq:L_SM_IR}
	\Lscr_\WCSM \bigr|_\IR & = 
	\sum_{i=1}^{12} \sum_{a=1}^2 \Psi_{ia}^\dagger i \bar{\sigma}^\mu \partial_\mu \Psi_{ia} 
	+ \sum_{i=1}^{21} \xi_i^\dagger i \bar{\sigma}^\mu \partial_\mu \xi_i 
	\\ & \quad 
	- \sum_{i,j=1}^{12} \sum_{a,b=1}^2 \bigl[ y_0 \LamW \varepsilon^{ab} \Psi_{ia} \Sigma_{ij}^\dagger \Psi_{jb} + \hc \bigr] 
	\nn & \quad 
	- \sum_{i,j=1}^{12} \sum_{a,b=1}^2 \bigl[ ( y_0^\prime \LamW / f) \varepsilon^{ab} \Psi_{ia} \Sigma_{ij}^\dagger \Psi_{jb} \Phi + \hc \bigr] 
	\nn & \quad 
	+ \frac{f^2}{4} \, \mathrm{Tr}\bigl[ \partial_\mu \Sigma^\dagger \partial^\mu \Sigma \bigr] 
	+ \kappa \LamW^2 f^2 {\rm Re}[{\rm det} \, \Sigma] 
	\nn & \quad 
	+ \frac{1}{2} \partial^\mu \Phi \partial_\mu \Phi 
	- \LamW^2 f^2 \, V\bigl(\Phi/f\bigr)
	 \nonumber
	\per
\end{align}
Recall that $y_0$, $y_0^\prime$, and $\kappa$ are $\mathcal{O}(1)$ dimensionless coefficients and $\LamW \sim 4 \pi f$ is the scale at which $\SU{2}_L$ weak isospin confines.  
This Lagrangian describes an effective field theory of the low-energy degrees of freedom where we have integrated out a tower of resonances with masses $m_\mathrm{res} \sim \LamW$, and we do not show all the non-renormalizable operators that they generate.  
The anomalous $\U{1}_\psi$ part of \eref{eq:WCSM_U12} is broken by the determinant term, for $\kappa \neq 0$, which arises from non-perturbative instanton effects in the confined phase.  

%=========
As we have discussed already in \sref{sec:toymodels}, confinement is accompanied by the spontaneous breaking of the $\U{12}_\psi = \SU{12}_\psi \times \U{1}_\psi$ global symmetry from \eref{eq:WCSM_U12}.  
The scalar condensate field $\Sigma(x)$ acquires a nonzero vacuum expectation value, $\Sigma_0 = \langle \Sigma \rangle$, and in order to achieve the anticipated symmetry breaking pattern, $\SU{12}_\psi / \Sp{12}_\psi$, we should have $\Sigma_0^\dagger \Sigma_0 = \bbone$~\cite{Lewis:2011zb,Arthur:2016dir,Haber}.  
As $\SU{12}_\psi$ has $143$ generators and $\Sp{12}_\psi$ has $78$ generators, then the number of broken generators is $65$, which equals the number of massless Goldstone boson fields, $\Pi^a(x)$.  
The vacuum expectation value $\Sigma_0 \neq 0$ also breaks the $\U{1}_\psi \subset \U{12}_\psi$, but since this symmetry is anomalous, the would-be Goldstone boson has its mass lifted by electroweak instantons, in analogy with the $\eta^\prime$ meson of QCD.  
Thus after confinement and spontaneous symmetry breaking, the symmetry group from \eref{eq:WCSM_U12} is broken to 
\begin{align}\label{eq:WCSM_SP12}
	\bigl\{ \Sp{12}_\psi \times \SU{2}_\phi \times \U{21}_\xi \bigr\}_\mathrm{global},
\end{align}
where $\Sp{12}_\psi$ acts (nonlinearly) on the $65$ ``pions,'' $\Pi^a$. 

%=========
Due to spontaneous symmetry breaking, $\Sigma_0 \neq 0$, the Yukawa interaction, $\sim \Psi \Sigma^\dagger \Psi$ in \eref{eq:L_SM_IR}, becomes a Dirac mass for the composite fermions.  
In particular, the $24$ composite fermions, $\Psi_{ia}$, form $12$ degenerate Dirac fermions with mass $m_\Psi \sim y_0 \LamW$, on the order of the confinement scale.  
Meanwhile the $21$ singlet fermions, $\xi_i$, remain massless and noninteracting to all orders, since the relevant operators are forbidden by the residual global symmetry \pref{eq:WCSM_SP12}.  

%=========
The composite scalar $\Phi$ has a potential $V$, which induces its mass and self-interaction.  
An analytical calculation of $V$ is hindered by the strongly-coupled nature of the theory, but dimensional analysis arguments suggest that $\Phi$ will acquire a nonzero vacuum expectation value, $\langle \Phi \rangle \sim f$.  
Since $\Phi$ is a singlet under the residual symmetry \pref{eq:WCSM_SP12} its vacuum expectation value does not signal any symmetry breaking.   

%=========
In summary, the spectrum of the WCSM (with $g_s = g^\prime = \lambda_\mathrm{Yuk} = 0$) in the confined phase is 
\begin{subequations}
\begin{align}
	m_\Pi = 0 & \qquad \text{65 Goldstone bosons} \\ 
	m_{\eta^\prime} \sim \LamW & \qquad \text{1 pseudo-Goldstone} \\ 
	m_{\Psi} \sim \LamW & \qquad \text{12 composite Dirac fermions} \\ 
	m_{\Phi} \sim \LamW & \qquad \text{1 composite scalar} \\ 
	m_{\xi} = 0 & \qquad \text{21 elementary Weyl fermions} \\ 
	m_\mathrm{res} \sim \LamW & \qquad \text{heavier resonances}\per
\end{align}
\end{subequations}
The heavy resonances are not included in \eref{eq:L_SM_IR}.  

%=========
Finally, we investigate the vacuum structure of this theory more carefully.  
We have seen that the symmetry breaking pattern $\SU{12}_\psi / \Sp{12}_\psi$ is obtained when the antisymmetric scalar condensate's vacuum expectation value satisfies $\Sigma_0^\dagger \Sigma_0 = \bbone$.  
To be more explicit, we work in a basis where 
\begin{align}\label{eq:Sigma0_WCSM}
	\Sigma_0 = \begin{pmatrix} 
	A &  &  \\ 
	& A & \\ 
	& & A 
	\end{pmatrix} 
	\, , \quad 
	A = \begin{pmatrix}
	0 & 1 & 0 & 0 \\ 
	-1 & 0 & 0 & 0 \\ 
	0 & 0 & 0 & 1 \\ 
	0 & 0 & -1 & 0
	\end{pmatrix} 
	\com
\end{align}
where the three blocks correspond to the three generations of fermions.  
Recall the mapping from the scalar condensate to the SM quark and lepton doublets, $\Sigma_{ij} \sim \psi_i \varepsilon \psi_j$ with $\psi_i$ given by \eref{eq:psi_and_chi}.  
Therefore the symmetry breaking pattern implies 
\begin{align}\label{eq:condensates}
	\langle l_i \varepsilon q_i^R \rangle \neq 0 
	\quad \text{and} \quad 
	\langle q_i^G \varepsilon q_i^B \rangle \neq 0
	\com
\end{align}
for each of the three generations $i \in \{ 1, 2, 3\}$.  
Thus we see that the vacuum state carries a nonzero baryon number and lepton number, and consequently the particles in \eref{eq:L_SM_IR} will experience $\mathsf{B}$- and $\mathsf{L}$-violating interactions that allow 
\begin{align} 
	3 \Delta \mathsf{B} = \Delta \mathsf{L} = \pm 1
	\quad \text{and} \quad 
	\Delta \mathsf{B} = \pm 2 / 3
	\per
\end{align}
The violation of baryon number is a requirement for any model of baryogenesis, and its natural emergence here is a tantalizing hint for a connection between the WCSM and the cosmological excess of matter over antimatter; we discuss this possibility further in \sref{sec:cosmoimplications}.  

%-----------------------
% Turn on strong, hypercharge, and Yukawa couplings
%-----------------------
\subsection{Turn on strong, hypercharge, and Yukawa couplings}\label{sec:turnoncouplings}

%=========
Let us finally consider $\SU{2}_L$ confinement with the Standard Model particle content at the measured values for the strong, hypercharge, and Yukawa couplings.  
Since these nonzero couplings explicitly break the large global flavor symmetry, we will find that the spectrum of pseudo-Goldstone pions is lifted.  
Moreover, because the condensates \pref{eq:condensates} spontaneously break color and hypercharge, we will see that some of the gluons and hypercharge boson acquire mass from the Higgs mechanism by ``eating'' the would-be Goldstone bosons.  
The reader may be familiar the style of analysis that we employ in this section from studies of Little Higgs theories~\cite{Perelstein:2005ka}.  

%=========
\paragraph*{Comparison with QCD.}  
Before beginning the determination of the effect of these explicit chiral symmetry breaking terms in the WCSM, we contrast the symmetry breaking to that in confining SM QCD.  The quark masses and electric couplings are both sources of explicit chiral symmetry breaking in the SM.  Since the quark masses only involve confining states, they contribute to chiral symmetry breaking terms at tree-level in the Lagrangian.  On the other hand, the electric coupling, which involves the non-confining photon, only contributes a small correction at loop level.  Furthermore, the same combination of the non-confining $\SU{2}_L \times \U{1}_Y$ gauge group that would get broken by confinement of QCD is already broken at a much higher scale by the Higgs mechanism, so all of the pions remain in the spectrum rather than being eaten.  On the other hand, there are no chiral symmetry breaking operators involving only the weak-charged fermions in the WCSM, so all chiral symmetry breaking for the pions occurs at loop level stemming from the $\SU{3}_c \times \U{1}_Y$ gauge and Yukawa couplings.  There is no higher scale breaking of the weakly-coupled $\SU{3}_c \times \U{1}_Y$ gauge group, so it is broken only by the confinement of $\SU{2}_L$, leading to both some eaten Goldstone modes and non-negligible corrections due to the massive gauge bosons that are not completely decoupled at the confinement scale.

%=========
\paragraph*{Treat small couplings perturbatively.}  
A key assumption in the following analysis is that the nonzero strong, hypercharge, and Yukawa couplings can be treated as perturbations on the preceding analysis of \sref{sec:turnoffcouplings}.  
This is a reasonable expectation as most of these couplings are $\ll \mathcal{O}(1)$, with the largest being the top Yukawa coupling and the strong coupling.  
Specifically, we assume that the scalar condensate, $\Sigma_{ij}$, still acquires a vacuum expectation value $\Sigma_0$ that leads to the $\SU{12}_\psi / \Sp{12}_\psi$ symmetry breaking pattern as in \eref{eq:Sigma_vev}.  

%=========
\paragraph*{Introduce gauge interactions.}  
We implement the $\SU{3}_c \times \U{1}_Y$ gauge interactions by promoting partial derivatives in the WCSM's IR Lagrangian \pref{eq:L_SM_IR} to covariant derivatives:  $D_\mu \Psi_{ia}$, $D_\mu \xi_i$, $D_\mu \Sigma_{ij}$, and $D_\mu \Phi$.  
Since the scalar $\Phi$ is a singlet under $\SU{3}_c \times \U{1}_Y$, we have simply $D_\mu \Phi = \partial_\mu \Phi$.  
Let us look more closely at $D_\mu \Sigma_{ij}$, as it reveals how the gauge boson masses arise through the Higgs mechanism.  
The covariant derivative of the scalar field $\Sigma_{ij}(x)$, is written as 
\begin{align}\label{eq:DSigma_WCSM}
%--
	D_\mu \Sigma & = \partial_\mu \Sigma - i g_s \sum_{a = 1}^8 G^a_\mu \bigl( L^a \Sigma + \Sigma L^{aT} \bigr)
	\\ & \qquad 
	- i g^\prime B_\mu \bigl( Y \Sigma + \Sigma Y \bigr), \nonumber 
%--
\end{align}
where $g_s$ is the $\SU{3}_c$ coupling, $G_\mu^a(x)$ for $a = 1, \cdots, 8$ are the gluon fields, $L^a$ are the generators of $\SU{3}_c$, $g^\prime$ is the $\U{1}_Y$ coupling, $B_\mu(x)$ is the hypercharge boson field, and $Y$ is the generator of $\U{1}_Y$.  
It is convenient to define the $\SU{3}_c$ generators in a slightly unconventional way using the following basis 
\begin{equation}
\begin{aligned}
\lambda^1 &= \dfrac{1}{2} \begin{pmatrix}
0 & 0 & 0 \\
0 & 0 & 1 \\
0 & 1 & 0 \end{pmatrix},
&&  \lambda^2  = \dfrac{1}{2} \begin{pmatrix}
0 & 0 & 0 \\
0 & 0 & -i \\
0 & i & 0 \end{pmatrix},  \\ 
 \lambda^3 &= \dfrac{1}{2} \begin{pmatrix}
0 & 0 & 0 \\
0 & 1 & 0 \\
0 & 0 & -1 \end{pmatrix}, 
 && \lambda^4 = \dfrac{1}{2} \begin{pmatrix}
0 & 1 & 0 \\
1 & 0 & 0 \\
0 & 0 & 0 \end{pmatrix} \\
 \lambda^5 & = \dfrac{1}{2} \begin{pmatrix}
0 & -i & 0 \\
i & 0 & 0 \\
0 & 0 & 0 \end{pmatrix}, 
&& \lambda^6 = \dfrac{1}{2} \begin{pmatrix}
0 & 0 & 1 \\
0 & 0 & 0 \\
1 & 0 & 0 \end{pmatrix}, \\ 
\lambda^7 &=\dfrac{1}{2} \begin{pmatrix}
0 & 0 & -i \\
0 & 0 & 0 \\
i & 0 & 0 \end{pmatrix}, 
&& \lambda^8 = \dfrac{1}{2\sqrt{3}}\begin{pmatrix}
-2 & 0 & 0 \\
0 & 1& 0 \\
0 & 0 &1 \end{pmatrix}
\per
\end{aligned}
\end{equation}
As such, the $\SU{3}_c$ and $\U{1}_Y$ generators are written as 
\begin{align}\label{eq:La_Y_def}
	L^a & = {\rm diag}(0, \lambda^a, 0, \lambda^a, 0, \lambda^a) 
	\quad \text{for} \ \ a = 1, \cdots, 8 \\ 
	Y & = {\rm diag}\Bigl(
-\frac{1}{2}, \frac{1}{6}, \frac{1}{6}, \frac{1}{6},
-\frac{1}{2}, \frac{1}{6}, \frac{1}{6}, \frac{1}{6},
-\frac{1}{2}, \frac{1}{6}, \frac{1}{6}, \frac{1}{6}
\Bigr) \nonumber
	\com
\end{align}
respectively.  
The generators are orthonormal such that $\mathrm{Tr}[L^a L^b] = 3 \delta^{ab}/2$, $\mathrm{Tr}[L^a Y] = 0$, and $\mathrm{Tr}[YY] = 1$.

%=========
\paragraph*{Explicit global symmetry breaking.}
By setting $g_s = g^\prime = \lambda_\mathrm{Yuk} = 0$ previously, we saw that the SM global symmetry of \eref{eq:SM_sym} was enhanced to \eref{eq:WCSM_U12}.  
Since the nonzero Yukawa couplings are typically much smaller than the $\SU{3}_c \times \U{1}_Y$ gauge couplings, it is also useful to consider the intermediate case in which $g_s, g^\prime \neq 0$ and $\lambda_\mathrm{Yuk} = 0$.  
For vanishing Yukawa couplings, the global symmetry group of the theory is 
\begin{align}\label{eq:global_no_Yukawa}
	& \SU{3}_q \times \SU{3}_U \times \SU{3}_D \times \SU{3}_l 
	\\ & \quad 
	\times \SU{3}_E \times \U{1}_U \times \U{1}_D \times \U{1}_E \times \U{1}_H 
	\com \nonumber
\end{align}
where each factor acts nontrivially on only one field, indicated by the subscript.  
In other words the nonzero gauge couplings explicitly break the global symmetry from \eref{eq:WCSM_U12} to \eref{eq:global_no_Yukawa}.  
This explicit breaking lifts several of the would-be Goldstone bosons, as we demonstrate below.  
The choice of the definition of the $\U{1}$ factors here is not unique; any linear combinations of these with $\U{1}_Y$ will also be a symmetry.  
We choose these combinations for later convenience.  
Two combinations of the residual $\U{1}$ global symmetry factors are anomalous under $\SU{3}_c \times \U{1}_Y$, but this anomalous explicit breaking is extremely weak and we neglect it in this work.

%=========
\paragraph*{Spontaneous symmetry breaking.}  
The nonzero vacuum expectation value, $\langle \Sigma \rangle = \Sigma_0$ in \eref{eq:Sigma0_WCSM}, indicates that the gauge symmetry is broken as follows:
\begin{equation}\label{eq:sym_breaking_with_couplings}
	 \{ \, \SU{3}_c \times \U{1}_Y \}_\mathrm{gauge} \to
	\{ \SU{2}_c \times \U{1}_Q \}_\mathrm{gauge} 
	\per 
\end{equation}
Specifically, $\SU{3}_c$ is broken to an $\SU{2}_c$ subgroup, and a linear combination of $\SU{3}_c$ and $\U{1}_Y$ is broken to a $\U{1}_Q$ subgroup.  
The generators of the $\SU{2}_c$ are simply the Pauli matrices, $\tau^a = \sigma^a/2$ for $a=1,2,3$, and the generator of $\U{1}_Q$ is 
\begin{align}\label{eq:Qgen}
	Q & = \frac{1}{\sqrt{3}} L^8 -Y \\ 
	& = {\rm diag}\Bigl(\frac{1}{2},-\frac{1}{2},0,0,\frac{1}{2},-\frac{1}{2},0,0,\frac{1}{2},-\frac{1}{2},0,0\Bigr) \nonumber
	\per
\end{align}
Note that $Q$ is not SM electromagnetism, but we use this notation to draw a parallel between \eref{eq:sym_breaking_with_couplings} and SM electroweak symmetry breaking.  
By counting the broken symmetry generators, we identify the gauge bosons that acquire mass through the Higgs mechanism.  
This will be discussed in detail shortly, but the result is that five gauge bosons acquire a mass and ``eat'' five of the would-be pions.

%=========
Nonzero $\Sigma_0$ also signals spontaneous breaking of the global symmetry acting on the $\SU{2}_L$-charged fermions.  
In the regime where the Yukawa couplings are neglected, we observe that the global symmetry of \eref{eq:global_no_Yukawa} is broken to 
\begin{align}\label{eq:many_syms}
	& \SO{3}_g \times \SU{3}_U \times \SU{3}_D \times \SU{3}_E 
	\\ & \quad 
	\times \U{1}_U \times \U{1}_D \times \U{1}_E \times \U{1}_H 
	\per \nonumber 
\end{align}
The $\SO{3}_g$ factor corresponds to a real rotation among the three generations.  
There are thus $13$ spontaneously broken global symmetry generators, and correspondingly we find $13$ massless Goldstone bosons remain after inclusion of the $\SU{3}_c \times \U{1}_Y$ gauge interactions, while the remaining $47$ pions are lifted at this level.	

%=========
\paragraph*{Charges under $\SU{2}_c \times \U{1}_Q$.}  
Let us take a brief aside to study how the residual gauge symmetry \pref{eq:sym_breaking_with_couplings} acts on the various fields.  
For the $\SU{2}_L$-doublet fermions, we find that $l_i \sim {\bm 1}$ for $i = 1,2,3$ transforms as a singlet under $\SU{2}_c$, while $q_i^R \sim {\bm 1}$ also transforms as a singlet and $(q_i^G , q_i^B) \sim {\bm 2}$ transforms as a doublet.  
To emphasize this distinction between the $\SU{2}_c$-singlet and doublet quark fields, we introduce the following notation: 
\begin{align}\label{eq:sym_breaking_with_couplings2}
	q_{Si}^{} = q_i^R 
	\qquad \text{and} \qquad 
	q_{Di}^{} = ( q_i^G \, , \, q_i^B ), 
\end{align}
where $i=1,2,3$ labels the generation.  
Additionally, \eref{eq:Qgen} reveals that $l_i$ and $q_{Si}$ carry charges $+1/2$ and $-1/2$, respectively, under $\U{1}_Q$, while $q_{Di}$ is not charged under $\U{1}_Q$.  
A similar decomposition is also possible for the $\SU{2}_L$-singlet quarks, $U_i$ and $D_i$, while the $\SU{2}_L$-singlet leptons, $E_i$, are singlets under the $\SU{2}_c$ and carry charge $-1$ under the $\U{1}_Q$.  
Finally, the Higgs doublet, as represented by $\varphi_a$ for $a=1,2$, can also be denoted as $\varphi^+ = \varphi_2$ and $\varphi^- = \varphi_1$, where the superscript indicates the sign of the $\U{1}_Q$ charge.  
These charge assignments are summarized in \tabref{tab:charge_assignments4}. 

%=========
\paragraph*{Composite scalars under $\SU{2}_c \times \U{1}_Q$.}  
We have seen that $\Sigma_{ij}(x)$ represents $66$ pseudoscalar fields: $\eta^\prime(x)$ and $\Pi^a(x)$ for $a = 1, \cdots, 65$.  
To clarify how these scalar fields transform under $\SU{2}_c \times \U{1}_Q$, it useful to map them to fermion bilinears as follows:  
\begin{subequations}\label{eq:pions}
\begin{align}
	\Pi_{\bm{1}}^+ & \sim \ l \varepsilon l - q_S^\dagger \varepsilon q_S^\dagger \\ 
	\Pi_{\bm{1}}^- & \sim \ q_S^{} \varepsilon q_S - l^\dagger \varepsilon l^\dagger \\ 
	\Pi_{\bm{1}}^0 & \sim \ l \varepsilon q_S^{} - q_S^{\dagger} \varepsilon l^\dagger 
	\ , \ \ q_D^{} \varepsilon \varepsilon_c q_D^{} - q_D^{\dagger} \varepsilon \varepsilon_c q_D^{\dagger} \\ 
	\Pi_{\bm{2}}^+ & \sim \ l \varepsilon q_D^{} - q_S^{\dagger} \varepsilon q_D^{\dagger} \\ 
	\Pi_{\bm{2}}^- & \sim \ q_S^{} \varepsilon q_D^{} - l^{\dagger} \varepsilon q_D^{\dagger} \\
	\Pi_{\bm{3}}^0 & \sim \ q_D^{} \varepsilon \sigma^a \, q_D^{} - q_D^{\dagger} \varepsilon \sigma^a \, q_D^{\dagger},
\end{align}
\end{subequations}
where $\varepsilon_c = i \sigma^2$ and $\sigma^a$ are the Pauli matrices with $\SU{2}_c$ indices, and there is an implicit contraction over $\SU{2}_L$ indices such that each $\Pi$ is an $\SU{2}_L$ singlet.  
Note that these $\Pi$'s correspond to a total of $66$ states, which includes the $\eta^\prime$, the $5$ Goldstone bosons eaten by the broken gauge fields and the $60$ would-be Goldstone bosons.\footnote{
We provide a few examples of counting these degrees of freedom.  For the $\Pi_{\bm{1}}^\pm$ there are three possible combinations of $i$ and $j$ as they must be anti-symmetric for each of $\pm$ which results in six $\Pi_{{\bm 1}}^\pm$ states (equivalently $N_{g}(N_{g}-1)$ where $N_{g}$ is the number of generations).  Likewise, for the $\Pi_{{\bm 3}}^0$ pion there are three combinations of generational indices each with three possible $SU(2)_c$ states which result in nine ($3N_{g}(N_{g}-1)/2$) pion states of $\Pi_{{\bm 3}}^0$.  }  

%=========
\begin{table}[t]
\begin{tabular}{|c||c|cc|c|}
\hline
&States & $\SU{2}_c$ & $\U{1}_Q$ & $\U{1}_{r}$ \\ 
\hline
$l_i$ & 6 & ${\bm 1}$ & 1/2 & 0 \\
$q_{Si}$ & 6 & ${\bm 1}$ & -1/2 & 0 \\
$q_{Di}$ & 12 & ${\bm 2}$ & 0 & 0   \\
$E_i$ & 3 & ${\bm 1}$ & -1 & -1\\
$U_{Si}$ & 3 & ${\bm 1}$ & 1 & 1\\ 
$U_{Di}$ & 6 & ${\bm 2}$ & 1/2 & 1\\
$D_{Si}$ & 3 & ${\bm 1}$ & 0 & -1\\
$D_{Di}$ & 6 & ${\bm 2}$ & -1/2 & -1\\ 
$\varphi^\pm$ & 4 & ${\bm 1}$ & $\pm$1/2 & $\pm 1$ \\
$G^{1,2,3}$ & 6 & ${\bm 3}$ & 0 & 0\\ 
$A^\prime$ & 2 & ${\bm 1}$ & 0 & 0\\ 
$W^{\pm \prime}$ & 12 & ${\bm 2}$ & $\pm1/2$ & 0\\ 
$Z^\prime$ & 3 & ${\bm 1}$ & 0 & 0\\ 
\hline\hline
$\Psi_{{\bm 1}}^0, \Psi_{{\bm 1}}^{0c}$ & 6 & ${\bm 1}$ & 0 & $\pm 1$ \\
$\Psi_{{\bm 1}}^{\pm}$ & 6 & ${\bm 1}$ & $\pm$1 & $\pm 1$ \\
$\Psi_{{\bm 2}}^{\pm}$& 12 & ${\bm 2}$ & $\pm$1/2 & $\pm 1$ \\
$\Xi_{{\bm 1}}^0$ & 3 & ${\bm 1}$ & 0 & -1 \\
$\Xi_{{\bm 1}}^{\pm}$ & 6 & ${\bm 1}$ & $\pm$1 & 1 \\
$\Xi_{{\bm 2}}^{\pm}$& 12 & ${\bm 2}$ & $\pm$1/2 & 1 \\
$\Pi_{\bm{1}}^\pm$ & 6 & ${\bm 1}$ & $\pm$1 & 0\\
$\Pi_{\bm{1}}^0$ & 13 & ${\bm 1}$ & 0 & 0\\
$\Pi_{\bm{2}}^\pm$ & 32 & ${\bm 2}$ & $\pm$1/2 & 0 \\  
$\Pi_{\bm{3}}^0$ & 9 & ${\bm 3}$ & 0 & 0 \\  
$\eta^\prime$ & 1 & ${\bm 1}$ & 0 & 0 \\  
$\Phi$ & 1 & ${\bm 1}$ & 0 & 0 \\
\hline
\end{tabular}
\caption{\label{tab:charge_assignments4}This table summarizes how the WCSM particle content transforms under the $\SU{2}_c \times \U{1}_Q$ subgroup of $\SU{3}_c \times \U{1}_Y$ and the global $\U{1}_r$.  Particles above the double bar correspond with states in \tabref{tab:charge_assignments2}, and those below the bar correspond to composite particles.  
The ``States'' column counts $1$ for each Weyl spinor degree of freedom and $1$ for each real boson. 
}
\end{table}

%=========
\paragraph*{Composite fermions under $\SU{2}_c \times \U{1}_Q$.}  
The composite fermion fields, $\Psi_{ia}(x)$, can also be decomposed into irreducible representations of the residual gauge symmetry.  
Using the same notation as above, the composite fermions \pref{eq:Phi_Psi} correspond to 
\begin{subequations}\label{eq:compositefermions}
\begin{align}
	\Psi_{{\bm 1}}^0 & \sim \ l \, \varphi^- \\ 
	\Psi_{{\bm 1}}^{+} & \sim \ l \, \varphi^+ \\ 
	\Psi_{{\bm 1}}^{-} & \sim \ q_S \, \varphi^- \\ 
	\Psi_{{\bm 1}}^{0c} & \sim \ q_S \, \varphi^+ \\ 
	\Psi_{{\bm 2}}^{-} & \sim \ q_D \, \varphi^- \\ 
	\Psi_{{\bm 2}}^{+} & \sim \ q_D \, \varphi^+ 
	\com 
\end{align}
\end{subequations}
and their charge assignments are shown in \tabref{tab:charge_assignments4}.  

%=========
\paragraph*{Massless gauge boson interactions.}  
We write the interactions of the four massless force carriers, corresponding to the unbroken symmetry group $\SU{2}_c \times \U{1}_Q$, as follows.  
The covariant derivative from \eref{eq:DSigma_WCSM} contains
\begin{align}\label{eq:DSigma_GA}
	D_\mu \Sigma & \supset - i g_s \sum_{a=1}^3 G_\mu^a \bigl( L^a \Sigma + \Sigma L^{aT} \bigr) 
	\\ & \qquad 
	- i e_Q A_\mu^\prime \bigl( Q \Sigma + \Sigma Q \bigr)
	\per \nonumber 
\end{align}
For the $\SU{2}_c$ interactions, $g_s$ is the $\SU{3}_c$ gauge coupling that also determines the $\SU{2}_c$ coupling strength, $G_\mu^{1,2,3}$ are the force carriers of $\SU{2}_c$, and $L^{1,2,3}$ from \eref{eq:La_Y_def} are the generators of $\SU{2}_c$.  
For the $\U{1}_Q$ interaction
\begin{equation}\label{eq:eS_def}
	e_Q = g^\prime \cos \theta_Q \approx g^\prime,
\end{equation}
is the gauge coupling and the mixing angle is given by
\begin{equation}\label{eq:thetaS_def}
	\sin\theta_Q = \frac{g^\prime}{\sqrt{3 g_s^2 + g^{\prime 2}}} 
\end{equation}
 which is $\theta_Q \approx g^\prime / \sqrt{3} g_s \ll 1$ in the regime of interest where $g^\prime \ll g_s$. The massless force carrier
 is 
\begin{equation}
	A_\mu^\prime = \cos \theta_Q \, G_\mu^8 + \sin \theta_Q \, B_\mu,
\end{equation}
 and $Q$ from \eref{eq:Qgen} is the generator of $\U{1}_Q$.  

%=========
\paragraph*{Massive gauge boson interactions.}  
The five remaining gauge bosons correspond to the broken symmetry generators of \eref{eq:sym_breaking_with_couplings}.  
These states can be arranged into a pair of complex vector fields $W_\mu^{\prime \pm}(x)$, and a single real vector field $Z_\mu^\prime(x)$.  
We use the notation $W^{\prime \pm}$ and $Z^\prime$ to highlight the analogy with the Standard Model weak gauge bosons, conventionally denoted by $W$ and $Z$.  
Their interactions are written as 
\begin{align}\label{eq:DSigma_WZ}
	D_\mu \Sigma & \supset - i \frac{g_s}{\sqrt{2}} \sum_{\pm} \sum_{i=1,2} W^{\prime i\pm}_\mu (L^{i\pm} \Sigma + \Sigma L^{i\mp})
	\\ & \qquad 
	- i \frac{e_Q}{s_Q c_Q} Z^\prime_{\mu} \bigl( J \Sigma + \Sigma J \bigr) \nonumber 
	\com
\end{align}
where we define the new notation in the remainder of this paragraph.  
The $\SU{3}_c$ factor has five broken generators, which are $L^{4,5,6,7}$, and the corresponding gauge fields, $G_\mu^{4,5,6,7}(x)$, pair up to form two complex, massive vector fields, $W_\mu^{\prime \pm}(x)$ corresponding to 
\begin{align}\label{eq:Lpm_def}
	L^{\pm} & = (L^4 \pm i L^5, L^6 \pm i L^7)
	\per
\end{align}
The interaction strength of the $W^{\prime \pm}$ vectors is inherited from the $\SU{3}_c$ gauge coupling, $g_s$.  
The one final broken generator is
\begin{align}\label{eq:Jgen}
	J = \frac{1}{\sqrt{3}} \, L^8 - s_Q^2 \, Q
	\com
\end{align}
where $s_Q = \sin \theta_Q$ and $c_Q = \cos \theta_Q$, and the corresponding massive vector field, $Z_\mu^\prime(x)$, can be written as
\begin{equation}
	Z_\mu^\prime = - \sin \theta_Q \, G_\mu^8 + \cos \theta_Q \, B_\mu 
	\per
\end{equation}
The $W^{\prime \pm}$ transform as a charged doublet under $\SU{2}_c \times \U{1}_Q$ while $Z^\prime$ is a singlet; see \tabref{tab:charge_assignments4}.  

%=========
\paragraph*{Gauge boson spectrum.}  
The scalar field $\Sigma_{ij}(x)$ acquires a nonzero vacuum expectation value, given by \eref{eq:Sigma0_WCSM}, and the Higgs mechanism splits the spectrum of gauge bosons.  
The mass terms can be derived from \erefs{eq:DSigma_GA}{eq:DSigma_WZ} by using $\Sigma = \Sigma_0$ from \eref{eq:Sigma0_WCSM}.  
The gauge boson spectrum is summarized as follows 
\begin{subequations}
\begin{align}
	m_{G^{1,2,3}} & = 0 \\ 
	m_{A^\prime} & = 0 \\ 
	m_{W^{\prime\pm}} & = \sqrt{3/2} \, g_s f \\ 
	m_{Z^\prime} & = \sqrt{2/3} \sqrt{3g_s^2 + g^{\prime 2}} f
	\per
\end{align}
\end{subequations}
The four massless vector fields, $G_\mu^{1,2,3}$ and $A_\mu^\prime$, are the force carriers of $\SU{2}_c \times \U{1}_Q$, while the two complex vectors $W_\mu^{\prime\pm}$ and the single vector $Z_\mu^\prime$ acquire mass through the Higgs mechanism.  

%=========
\paragraph*{Symmetry breaking including Yukawa interactions.}
Finally we  discuss the ultimate spectrum of pions.  
Recall that for vanishing strong, hypercharge, and Yukawa couplings, the scalar $\Sigma_{ij}(x)$ encodes $65$ massless Goldstone bosons $\Pi^a(x)$, which we called ``pions,'' and one massive pseudoscalar $\eta^\prime(x)$, which is associated with an anomalous $\U{1}_\psi$ and acquires mass from $\SU{2}_L$ instantons.  
Taking the strong, hypercharge, and Yukawa couplings to be nonzero breaks the large global symmetry of \eref{eq:WCSM_U12} down to the $\U{1}_{\Bsf/3-\Lsf_i}^3$ of \eref{eq:SM_sym}.  

Thus we expect that all but at most three of the pions will have their masses lifted by the effect of nonzero $g_s$, $g^\prime$, and $\lambda_\mathrm{Yuk}$.  
Since the gauge groups $\SU{3}_c$ and $\U{1}_Y$ are subgroups of $\SU{12}_\psi \times \SU{21}_\xi$, the gauge generators are spurions of $\SU{12}_\psi$ transforming as adjoints.  
The Yukawas are spurions transforming as anti-fundamentals of $\SU{12}_\psi$, as well as fundamentals under $\SU{12}_\xi$ as 
 summarized in \tref{tab:charge_assignments3}.  

%=========
The global symmetry is also spontaneously broken, but there is an unbroken $\U{1}_r$ symmetry acting only on the weak singlet fermions and Higgs in the UV or, correspondingly, only on the fermion states in the IR.  The charge can be written as the following combination of $\U{1}$ charges
\begin{equation}\label{eq:r_def}
r = -2 Y + \sum_i \left(\frac{\Bsf}{3} - \Lsf_i \right) = -2 Y + \Bsf - \Lsf \per
\end{equation}
This symmetry is an exact global symmetry of the Lagrangian and is not anomalous under the surviving gauge symmetries in the IR.  
It has important consequences for the fermion spectrum and baryogenesis.  
The charges of the spectrum under it are shown in Table \ref{tab:charge_assignments4}.  
There are thus two broken global symmetry generators and, correspondingly, two massless pions in the WCSM.  
The 11 other pions that were massless after inclusion of the $\SU{3}_c \times \U{1}_Y$ gauge interactions (see below \eref{eq:many_syms}) get lifted, such that a total of 58 of the 60 physical pions are massive.  

%=========
\begin{table}
\begin{tabular}{|c||c||c|c|c|c|c|}
\hline
& \multicolumn{2}{|c|}{$\SO{4}$} & \multicolumn{2}{|c|}{$\U{12}_\psi$} & \multicolumn{2}{|c|}{$\U{21}_\xi$} \\
\cline{2-7}
& $\SU{2}_L$ & $\SU{2}_\phi$ & \multicolumn{1}{|c|}{$\SU{12}_\psi$} & $\U{1}_\psi$ & $\SU{21}_\xi$ & $\U{1}_\xi$ \\ \hline 
$T^a,Y$ & {$\bm 1$} & {$\bm 1$} & \multicolumn{1}{|c|}{$\bm{143}$} & 0 & {$\bm 1$} & $0$ \\ 
$\lambda_u, \lambda_d, \lambda_e$ & {$\bm 1$} & {$\bm 2$} & \multicolumn{1}{|c|}{$\overline{\bm{12}}$} & -$q_\psi$ & {$\overline{\bm{21}}$} & -$q_\xi$ \\  \hline
\end{tabular}
\caption{\label{tab:charge_assignments3}
In the spurion analysis of \secref{sec:turnoncouplings}, this table shows the charge assignments of the $\SU{3}_c$ generators $T^a$, the $\U{1}_Y$ generator $Y$, and the Yukawa couplings $\lambda_u$, $\lambda_d$, and $\lambda_e$.  
 }
\end{table}

%=========
\paragraph*{Gauge corrections to pion masses.}
The strong and hypercharge gauge interactions (for $g_s \neq 0$ and $g^\prime \neq 0$) induce radiative corrections that split the pion mass spectrum.  
Working in the mass basis of the vector bosons, where the leading contribution for each gauge boson eigenstate sums up without mixing, the relevant mass-correction operators are written as~\cite{Coleman:1973jx}
\begin{align}\label{eq:DeltaL_pion_gauge}
	\Delta \Lscr & = 
	C_G \LamW^2 f^2 \frac{g_s^2}{16\pi^2} \sum_{a=1,2,3} \mathrm{Tr}[L^{aT} \Sigma^\dagger L^a \Sigma] 
	\\ & \quad 
	+ C_A \LamW^2 f^2 \frac{e_Q^2}{16\pi^2} \mathrm{Tr}[Q \Sigma^\dagger Q \Sigma] 
	\nn & \quad 
	+ C_W \LamW^2 f^2 \frac{g_s^2/2}{16\pi^2} \sum_\pm \sum_{i=1,2} \mathrm{Tr}[L^{i\pm} \Sigma^\dagger L^{i\pm} \Sigma] 
	\nn & \quad 
	+ C_Z \LamW^2 f^2 \frac{e_Q^2 / s_Q^2 c_Q^2}{16\pi^2} \ \mathrm{Tr}[J \Sigma^\dagger J \Sigma] 
	\com \nonumber
\end{align}
where $e_Q \approx g^\prime$ was given by \eref{eq:eS_def}, $L^a$ was given by \eref{eq:La_Y_def}, $Q$ was given by \eref{eq:Qgen}, $L^\pm$ was given by \eref{eq:Lpm_def}, $J$ was given by \eref{eq:Jgen}, and $\LamW \approx 4 \pi f$.  
Each term corresponds to a one-loop correction to the pion's self-energy; the first term arises from a $G^{1,2,3}$ loop, the second from an $A^\prime$ loop, the third from a $W^{\prime\pm}$ loop, and the fourth term arises from a $Z^\prime$ loop.  
The dimensionless operator coefficients -- $C_G$, $C_A$, $C_W$, and $C_Z$ -- encode the details of the loop calculation, and they are expected to be $\mathcal{O}(1)$ numbers~\cite{Das:1967it,Ayyar:2019exp}
with $C_A = C_G$ since these two operators both correspond to massless vectors in the loop.   

%=========
\paragraph*{Yukawa corrections to pion masses.}
Yukawa interactions between the pions and fermions (for $\lambda_\mathrm{Yuk} \neq 0$) also contribute to the pion's self-energy at one-loop order.  
The leading non-trivial contributions due to the Yukawas come in at fourth power, taking the form 
\begin{equation}\label{eq:DeltaL_pion_Yuk}
	\Delta \Lscr = C_{\mathrm{Yuk}} \frac{\LamW^2 f^2}{16\pi^2} \sum_{k,k^\prime} \sum_{a,b=1}^2 \mathrm{Tr}[(\lambda_{k a} \lambda_{k a}^\dagger) \Sigma^\dagger (\lambda_{k^\prime b}^\ast \lambda_{k^\prime b}^T) \Sigma]
	\com
\end{equation}
where $\lambda_u$ is the up-type quark Yukawa matrix, $\lambda_d$ is the down-type quark Yukawa matrix, $\lambda_e$ is the charged lepton Yukawa matrix and the indices $k,k^\prime$ run over $u,d,e$.  
These are $12 \times 12$ complex matrices in the basis of \eref{eq:psi_and_chi}.  

%=========
\paragraph*{Pion mass spectrum.}
\begin{table}
\begin{tabular}{c c c}
	\hline\hline
	Pion & \# & Squared Mass $ \times 16 \pi^2 / \LamW^2$ \\
	\hline

	\multirow{2}{*}{$\Pi_{\mathbf{3}}^0$} 
	& 3 & $-4C_G g_s^2$ \\ 
	& 6 & $- 4 C_G g_s^2 + 2 C_{\rm Yuk} |V_{tb}|^4 y_t^4$ \\ \hline 

	\multirow{6}{*}{$\Pi_{\mathbf{2}}^{\pm}$} 
	& \multirow{2}{*}{8} & $ - \frac{3}{2} C_G g_s^2 - \frac{1}{2} C_A e_Q^2 + \frac{1}{2} C_Z g_s^2 + C_{\rm Yuk} |V_{tb}|^4 y_t^4$ \\ 
	& & $- \sqrt{C_W^2 g_s^4 + C_\mathrm{Yuk}^2 |V_{tb}|^8 y_t^8} $ \\ 
	& \multirow{2}{*}{8} & $ - \frac{3}{2} C_G g_s^2 - \frac{1}{2} C_A e_Q^2 + \frac{1}{2} C_Z g_s^2 + C_{\rm Yuk} |V_{tb}|^4 y_t^4$ \\ 
	& & $+ \sqrt{C_W^2 g_s^4 + C_\mathrm{Yuk}^2 |V_{tb}|^8 y_t^8} $ \\ 
	& 12 & $ - \frac{3}{2} C_G g_s^2 - \frac{1}{2} C_A e_Q^2 + \frac{1}{2} C_Z g_s^2 + C_W g_s^2$ \\ 
	& 4 & $ - \frac{3}{2} C_G g_s^2 - \frac{1}{2} C_A e_Q^2  + \frac{1}{2} C_Z g_s^2 - C_W g_s^2$ \\ \hline 
	
	$\Pi_{\mathbf{1}}^{\pm}$ 
	& 6 & $ - 2 C_A e_Q^2 - \frac{2}{3} C_Z g_s^2$ \\ \hline 
	
	\multirow{8}{*}{$\Pi_{\mathbf{1}}^{0}$} 
	& 2 & $2 C_{\rm Yuk} |V_{tb}|^2 y_t^4$ \\
	& 4 & $2 C_{\rm Yuk} |V_{tb}|^2 y_t^2 y_\tau^2$ \\
	& 1 & $2 C_{\rm Yuk} |V_{cs}|^4 y_c^4$ \\
	& 2 & $- \frac{1}{2} C_{\rm Yuk} {\rm Re}[V_{ts}]^2 y_t^2 y_\tau^2$ \\
	& 1 & $- 2 C_{\rm Yuk} {\rm Re}[V_{ts}]^2 y_t^2 y_\tau^2$ \\
	& 1 & $6 C_{\rm Yuk} {\rm Re}[V_{cs}] {\rm Re}[V_{td}] \frac{{\rm Re}[V_{cs}]{\rm Re}[V_{td}] - {\rm Re}[V_{cd}]{\rm Re}[V_{ts}]}{{\rm Re}[V_{ts}]^2} y_c^2 y_\mu^2$ \\
	& 2 & 0 \\ 
	
	\hline\hline
\end{tabular}
\caption{Approximate mass spectrum for the pions.  To simplify the expressions we assume that all the $C$'s are $O(1)$ and show only the leading contributions.  The second column indicates the multiplicity of each state, although some degeneracies are split by higher order contributions.  Five would-be pions, which are ``eaten'' by the massive $W^{\prime \pm}$ and $Z^\prime$ bosons, are not shown in the table.  
}\label{tab:pion-spectrum}
\end{table}
To calculate the pion mass spectrum, we parametrize the pions $\Pi^a$ in terms of $\Sigma_{ij}$ using \eref{eq:Sigma_def}, evaluate \erefs{eq:DeltaL_pion_gauge}{eq:DeltaL_pion_Yuk}, and read off the quadratic terms, $\Delta \Lscr = - (1/2) (M^2_\Pi)_{ab} \Pi^a \Pi^b$.  The leading masses of the pions are shown in Table \ref{tab:pion-spectrum}.  Including all of the mass-correction operators, the pion spectrum contains only two massless states, corresponding to the Goldstone bosons associated with $\U{1}_Y \times \U{1}_{\Bsf/3-\Lsf_i}^3 / \U{1}_Q \times \U{1}_r$.  If all $C$'s are nonzero then one cannot avoid some tachyonic states, but choosing $C_G < 0$, $C_A < 0$, $1 \lesssim C_Z < 0$, $|C_W| \lesssim 1$, and $C_\mathrm{Yuk} > 0$ leads to only four slightly tachyonic $\Pi_{\bm 1}^0$ with masses suppressed by the small Yukawas $y_\tau$ and $y_c$.  This indicates that the assumed symmetry breaking pattern, $\SU{2N_f} / \Sp{2N_f}$, is radiatively stable modulo small corrections.  

%=========
\paragraph*{Fermion mass spectrum}
As we have discussed in \secref{sec:turnoffcouplings}, the composite fermions $\Psi_{ia}$ get a mass of order $\LamW$ from the confinement process.  
However, once we include the Yukawa interactions, the composite fermions acquire a small mixing with the $\SU{2}_L$-singlet fermions, $\xi_j$, and the spectrum is split.  
To see this explicitly, we construct the Lagrangian for the fermions.  
In the UV the Yukawa interactions take the form 
\begin{align}\label{eq:LYuk_UV}
	\Delta \Lscr_\mathrm{Yuk} \bigr|_\UV 
	& = \sum_{i=1}^{12} \sum_{j=1}^{21} \sum_{a=1}^2 
	(\lambda_u + \lambda_d + \lambda_e)_{ija} 
	\varphi_a \psi_i \xi_j 
	\per
\end{align}
In the IR the $\SU{2}_L$-doublet fermions are confined with the Higgs, and this Lagrangian is mapped to
\begin{align}\label{eq:LYuk_IR}
	\Delta\Lscr_{\mathrm{Yuk}} \bigr|_\IR 
	& = C_\lambda f \sum_{i=1}^{12} \sum_{j=1}^{21} \sum_{a=1}^2 
	(\lambda_u + \lambda_d + \lambda_e)_{ija} \Psi_{ia} \xi_j
	\com
\end{align}
where $C_\lambda$ is an $\mathcal{O}(1)$ constant.  
This operator allows the composite $\Psi_{ia}$ to mix with the elementary $\xi_j$, and the resultant mass eigenstates are mostly-composite fermions with masses $\sim \LamW$ and mostly-elementary fermions, which we denote by $\Xi$, with masses $\ll \LamW$.  
The mostly-elementary fermion mass eigenstates correspond to the following elementary constituents:
\begin{subequations}
	\begin{align}
	\Xi^0_{\bm 1} & \sim D_{S} \\ 
	\Xi^+_{\bm 1} & \sim U_{S} \\ 
	\Xi^-_{\bm 1} & \sim E \\ 
	\Xi^+_{\bm 2} & \sim U_{D} \\ 
	\Xi^-_{\bm 2} & \sim D_{D} 
	\per
	\end{align}
\end{subequations}
The masses for these fermions arise only in the presence of non-zero Yukawa couplings.  
The charged fermions get Dirac masses via a seesaw mechanism as follows
\begin{subequations}\label{eq:composite-masses}
\begin{align}
	m_{\Xi^\pm_{\bm 1}} & \approx C_\lambda^2 \lambda_e \lambda_u \frac{f^2}{y_0 \LamW} \\ 
	m_{\Xi^\pm_{\bm 2}} & \approx C_\lambda^2 \lambda_u \lambda_d \frac{f^2}{y_0 \LamW}
	\per
\end{align}
\end{subequations}
Note that the CKM matrix should appear in this mass matrix, but can be rotated to generate an off-diagonal $W^\prime$ coupling, as in the SM.  The mass of each of generation of composite fermions is then determined by the product diagonal Yukawa matrices of that generation as in \eqref{eq:composite-masses}.  The mostly composite heavy fermions also get some very small correction due to the Yukawa couplings.  The $\Xi^0_{\bm 1}$ fermions are protected from getting a mass by the $\U{1}_{r}$ symmetry and remain exactly massless Weyl fermions.  The light, mostly-elementary fermions mix with the heavy, mostly-composite fermions with a mixing angle given by
\begin{equation}
	\theta_{\mathrm{LH}} \sim \frac{C_\lambda \lambda_{u,d,e} f}{y_0 \LamW}
	\com
\end{equation}
and the mass eigenstates are partially composite~\cite{Kaplan:1991dc}.  
Since the Yukawas are typically $\lambda \ll 1$, this is a small mixing.  

%-----------------------
% Summary of the WCSM
%-----------------------
\subsection{Summary of the WCSM}\label{sec:WCSM_summary}

%=========
We have studied the symmetries and the lightest states of the WCSM.  After all interactions and spontaneous symmmetry breaking, the symmetry group of the WCSM is
\begin{equation}
\bigl\{\SU{2}_c \times \U{1}_Q\bigr\}_\mathrm{gauge} \times \bigl\{\U{1}_r\bigr\}_\mathrm{global}\per
\end{equation}
The lightest states in the spectrum include:
\begin{itemize}
\itemsep0em
	\item 2 massless Goldstone bosons corresponding to 2 combinations of spontaneous $\Bsf/3 - \Lsf_i$ breaking;
	\item 3 mostly-elementary massless Weyl fermions corresponding to the $\SU{2}_c \times \U{1}_Q$ singlets charged under $\U{1}_r$. They are mostly $\SU{2}_L$-singlet down-type fermions;
	\item 4 massless gauge bosons of $\SU{2}_c \times \U{1}_Q$.
	\item 47 pseudo-Goldstone bosons of the spontaneous breaking of $\SU{12}_\psi/\Sp{12}_\psi$  are dominantly  lifted by the $\SU{2}_c \times \U{1}_Q$ gauge interactions. The gauge interactions explicitly break the global symmetry $\SU{12}_\psi \times \SU{2}_\phi \times \U{21}_\xi/\SU{3}^5 \times \U{1}^4$, which gets spontaneously broken in the confined phase to a global symmetry $\SO{3}_g \times \U{3}^3 \times \U{1}$, leaving 13 massless Goldstone bosons after their inclusion;
	\item 11 pseudo-Goldstone bosons of the spontaneous breaking of $\SU{12}_\psi/\Sp{12}_\psi$ get lifted by the Yukawa interactions only. The Yukawas explicitly break $\SU{3}^5 \times \U{1}^4/ \U{1}_{\Bsf/3 - \Lsf_i}^3$, which gets spontaneously broken to $\U{1}_r$.  There are only 2 massless Goldstones bosons after their inclusion;
	\item 9 mostly-elementary Dirac fermions that get lifted by the Yukawa couplings.  They correspond mostly to the remaining weak-singlet fermions;
	\item 5 massive gauge bosons of the broken $\SU{2}_c \times \U{1}_Q$ gauge group.  These eat 5 of the would-be pions;
	\item 1 $\eta'$ pion corresponding to the anomalous $\U{1}_{\Bsf + \Lsf}$ global symmetry and acquires a mass of order $\LamW$;
	\item 1 Higgs-Higgs composite scalar that is not protected by any symmetry and gets a mass of order $\LamW$;
	\item 12 Dirac fermions corresponding mostly to fermion-Higgs composite states and they receive Dirac masses of order $\LamW$;
	\item Additional scalar, radial, and spin excitations that are not included in the non-linear sigma model.
\end{itemize}

%==================
% Potential Cosmological Implications
%==================
\section{Potential Cosmological Implications}\label{sec:cosmoimplications}

%=========
How would we know if the Universe passed through a phase of confined $\SU{2}_L$?  
In this section we briefly discuss the possible implications for various cosmological relics.  

%=========
\paragraph*{Cosmological phase transition.}  
After cosmological inflation evacuates the Universe of matter, we suppose that reheating produces a plasma of Standard Model particles at temperatures above the scale of weak confinement.  
As the Universe grows older and expands, the cosmological plasma cools to the confinement scale, and the confinement of the weak force is accomplished through a cosmological phase transition. 
This transition is expected to be first order, meaning that bubbles of the confined phase nucleate in a background of the unconfined phase, and the transition completes as the bubbles grow and merge, filling all of space.  
The first order nature of the weak-confining phase transition follows from a more general argument by Pisarski and Wilczek (1983)~\cite{Pisarski:1983ms}, who considered a general Yang-Mills gauge theory with $N_f$ flavors of massless vector-like fermions, and showed that it will confine through a first order phase transition if $N_f \geq 3$.  
As we have discussed already in \sref{sec:WCSM}, the Standard Model corresponds to $2N_f = 12$ massless flavors of $\SU{2}_L$ doublets, and therefore we expect the weak-confining phase transition to be first order.  
A first order cosmological phase transition typically provides the right environment for the creation of cosmological relics, such as topological defects, gravitational wave radiation, and the matter-antimatter asymmetry.  

%=========
\paragraph*{Topological defects.}  
When spontaneous symmetry breaking is accomplished through a cosmological phase transition, causality arguments require the symmetry-breaking scalar field to be inhomogeneous on the Hubble scale, and this inhomogeneity may allow the field to form topological defects, such as cosmic strings and domain walls~\cite{Kibble:1976sj}.  
In the context of QCD's chiral phase transition, where the $\SU{3}_A \times \U{1}_A$ symmetry is spontaneously broken by the chiral quark condensate, these defects have been called pion and eta-prime strings~\cite{Zhang:1997is}. As the nonzero quark masses break the $\SU{3}_A \times \U{1}_A$ symmetry explicitly, this may prevent the QCD defects from forming at all, but if they do form, then they are expected to be unstable and decay soon after the QCD phase transition~\cite{Eto:2013bxa}.  
Similarly we expect that spontaneous symmetry breaking in the WCSM will generate a string-wall defect network with a rich structure.  
For instance, spontaneously breaking the global symmetry $\U{1}_{\Bsf/3-\Lsf_i}^3$ in \eref{eq:SM_sym} is expected to generate multiple networks of stable cosmic strings, which could survive until $\SU{2}_L$ is de-confined.  
While the defect network is present, it may play a role in primordial magnetogenesis~\cite{Brandenberger:1998ew} or baryogenesis~\cite{Brandenberger:1992ys}.  

%=========
\paragraph*{Gravitational wave radiation.}  
A first order cosmological phase transition produces a stochastic background of gravitational waves~\cite{Hogan:1986qda} due to the collisions of bubbles and their interactions with the plasma.  
For a weakly-coupled theory, it is straightforward to calculate the spectrum of this gravitational wave radiation~\cite{Kamionkowski:1993fg}, and gravitational waves have been studied in strongly-coupled theories as well~\cite{Helmboldt:2019pan}.  
Since gravity is extremely weakly interacting, the gravitational wave radiation is expected to free stream to us today where it might be detected.  
For instance, gravitational waves produced at an electroweak-scale phase transition could be detected by space-based gravitational wave interferometer experiments like LISA~\cite{Caprini:2015zlo}.  
Measuring the appropriate spectrum of gravitational wave radiation would provide one of the most direct probes of weak confinement in the early Universe.  

%=========
\paragraph*{The matter-antimatter asymmetry.}  
In general a first-order cosmological phase transition also provides the right environment to explain the cosmological excess of matter over antimatter through the physics of baryogenesis~\cite{Kolb:1990}.   
At the weak-confining phase transition in particular, weak sphalerons are active in front of the bubble wall where they mediate baryon-number-violating reactions, and they become suppressed behind the bubble wall where $\SU{2}_L$ is confined and the weak-magnetic fields are screened.  
This observation suggests that it may be possible to generate a baryon asymmetry at the confined-phase bubble walls through dynamics similar to electroweak baryogenesis~\cite{Cohen:1993nk} or spontaneous baryogenesis~\cite{Kuzmin:1992up,Servant:2014bla,Ipek:2018lhm}.  

%=========
However, although sphalerons may be suppressed in the confined phase, this does not imply that $\Bsf + \Lsf$ is effectively conserved, as we find in the SM.  
As discussed in \sref{sec:WCSM}, baryon number is not a good quantum number in the confined phase where the quarks and leptons condense and acquire a vacuum expectation value, spontaneously breaking $\U{1}_{\Bsf+L} \times \U{1}_{\Bsf/3-\Lsf_i}^3$.  
Instead there is only one conserved global charge in the confined phase, $r = -2Y + (\Bsf-\Lsf)$ from \eref{eq:r_def}, which is not  anomalous, and consequently an asymmetry in $r$ cannot be generated by the weak sphaleron.  
All other global charges are violated by interactions in the Lagrangian or by the quark-lepton condensates.  
Since an asymmetry in a non-conserved global charge can be washed out in the confined phase, this observation suggests that WCSM does not have all the necessary ingredients for baryogenesis during the confining phase transition.  

%=========
Nevertheless, one can easily imagine extending the WCSM by heavy Majorana neutrinos so as to accommodate the observed neutrino masses and mixings.  
The associated lepton-number violation could provide a means of generating $r$-charge during the confining phase transition, which is later distributed into baryon number when the system de-confines and enters the standard low-temperature Higgs phase.  

%=========
Alternatively, it may be possible to accomplish baryogenesis during the deconfining phase transition.  
For this scenario to work, the system must pass directly from the confined phase into the Higgs phase where the sphalerons are inactive; if the system passes instead through the Coulomb phase then the sphalerons will come back into equilibrium and wash out any baryon asymmetry.  
This constraint has the interesting implication that the system must remain in the confined phase even at a time where the cosmological plasma temperature is quite low, $\lesssim 100 \GeV$.  
If the weak-confined phase results from the nontrivial dynamics of a modulus field, such as we discussed in the model of \eref{eq:L_modulus}, then this constraint implies a low mass for the modulus field.  

%==================
% Comparison with the Abbott-Farhi Model
%==================
\section{Comparison with the Abbott-Farhi Model}\label{sec:AbbottFarhi}

%=========
We  explored the idea that the $\SU{2}_L$ weak force may have been confined in the early Universe, only to become weakly coupled and Higgsed at later times.  
Confinement of the weak force was first studied in the pioneering work by Abbot and Farhi~\cite{Abbott:1981re,Abbott:1981yg}, which developed into a theory~\cite{Claudson:1986ch,tHooft:1998ifg,Calmet:2000th} that we will call the Abbott-Farhi Model (AFM).   
The AFM describes the SM particle spectrum as composite states that arise from the confinement of the $\SU{2}_L$ weak force at a scale $\LamW = \sqrt{G_F} \sim 300 \GeV$.  
These ideas were motivated in part by the complementarity between a Yang-Mills gauge theory in the Higgs phase and the confined phase~\cite{Osterwalder:1977pc,Fradkin:1978dv,Banks:1979fi,Dimopoulos:1980hn} (see also Refs.~\cite{Grady:2015dsa,Seiler:2015rwa}).
While the AFM was able to reproduce the observed mass spectrum, interactions, and even weak boson phenomenology, it was eventually found to be in tension with electroweak precision measurements at the $Z$-pole~\cite{Sather:1995tw,DEramo:2009eqs}, leaving the (Higgs-phase) SM as the preferred description of weak-scale phenomenology.  

%=========
A key assumption of the AFM, and the essential distinction with our work here, has to do with the assumed pattern of symmetry breaking.  
The AFM assumes that the chiral symmetry remains unbroken after confinement (in the regime where the hypercharge, QCD, and Yukawa interactions can be neglected).  
In our notation, this assumption corresponds to $\langle \Sigma \rangle = 0$.  
If the chiral symmetry is unbroken, then the AFM's composite fermions, corresponding to the familiar SM particle content, must be massless.  
In this way, the AFM's fermion masses are induced by the nonzero hypercharge, QCD, and Yukawa interactions, which explains why these particles are much lighter than the confinement scale $\LamW \sim 300 \GeV$.  
By contrast, we are motivated by recent numerical lattice studies to assume an $\SU{2N_f} / \Sp{2N_f}$ pattern of chiral symmetry breaking \pref{eq:Sigma_vev}, which leads to a massive spectrum of composite fermions with $m_\Psi \sim \LamW$.  

%=========
Which of these two symmetry-breaking patterns is the ``correct'' one?  
Since the physics controlling symmetry breaking is inherently non-perturbative, it is prohibitive to determine the symmetry breaking pattern through analytical techniques.  
Arguably, the assumed absence of chiral symmetry breaking that underlies the AFM did not follow from any fundamental physical principle, but rather it was taken as a desirable assumption in order to reproduced the known particle mass spectrum.  
However, recent numerical lattice studies~\cite{Lewis:2011zb,Arthur:2016dir} have provided evidence against the AFM's assumed absence of symmetry breaking and in favor of an $\SU{2N} / \Sp{2N}$ symmetry-breaking pattern instead.  
Motivated by these lattice results, we adopt the $\SU{2N} / \Sp{2N}$ breaking in our work \pref{eq:Sigma_vev}, and all of our results for the spectrum and interactions of composite particles in the early-universe weak-confined phase are predicated on this assumption.  

%==================
% Summary
%==================
\section{Summary}\label{sec:Summary}

%=========
We have studied a scenario in which the weak force is strong and the strong force is weak.  
In other words, we assume that new physics allows the $\SU{2}_L$ weak force to become strongly coupled and confine at a scale $\LamW > \mathrm{TeV}$ where the QCD strong force is weakly coupled, and we refer to this scenario as the weak-confined Standard Model (WCSM).  
We describe how interactions between the Standard Model and new physics may allow the WCSM phase to be realized in the early Universe at a time when the cosmological plasma temperature was sufficiently high, $T \sim \LamW$.  
By assuming a pattern of chiral symmetry breaking that it motivated by numerical lattice studies, we calculate the spectrum of low-lying composite particles, discuss their interactions, and compare this exotic phase with strong color confinement.  
Our analysis of the WCSM differs notably from earlier work on weak confinement, where it was assumed that confinement did not lead to chiral symmetry breaking.  
Although direct tests of the weak-confined phase will be challenging, given the richness of the WCSM spectrum there are several possible cosmological implications of this exotic phase in the early Universe.  
These cosmological observables, including topological defects, gravitational wave radiation, and baryogenesis, may provide fruitful avenues for further study.  

%==================
% Acknowledgments
%==================
\begin{acknowledgments}
We are grateful to Seyda Ipek whose thought-provoking colloquium at the Aspen Center for Physics stimulated this work.  
We would also like to thank Bogdan Dobrescu, Nima Arkani-Hamed, Anson Hook, William Jay, Aaron Pierce, Chris Quigg, Tim Tait, and Jesse Thaler for valuable discussions and encouragement.  
J.B.~is supported by PITT PACC.
A.J.L. is supported at the University of Michigan by the US Department of Energy under grant DE-SC0007859.
This manuscript has been authored by Fermi Research Alliance, LLC under Contract No. DE-AC02-07CH11359 with the U.S. Department of Energy, Office of Science, Office of High Energy Physics. 
This work was initiated and completed at the Aspen Center for Physics, which is supported by National Science Foundation grant PHY-1607611.
\end{acknowledgments}

%==================
% Bibliography
%==================
\bibliographystyle{apsrev4-1}
\bibliography{confined_SU2L}{}

\end{document}